\documentclass[aps,pra,preprint,superscriptaddress,amsmath,showpacs,tightenlines,nofootinbib]{revtex4-1}

\usepackage[ngerman,english]{babel}
\makeatletter\AtBeginDocument{\let\@elt\relax}\makeatother
\usepackage{amsmath}
\usepackage{slashed}
\usepackage[utf8]{inputenc}
\usepackage{fontenc}
\usepackage{braket}
\usepackage{bbm}
\usepackage{cancel}
\usepackage{graphicx}
\usepackage[outdir=./]{epstopdf}
\usepackage{calrsfs}
\usepackage{emptypage}
\usepackage{fancyhdr}

\usepackage[pdfstartview=FitB,colorlinks=false]{hyperref}
\hypersetup{
	pdftitle={Halo EFT for Ne-31 in spherical coordinates},
	pdfauthor={Wael Elkamhawy},
	pdfview=FitB
	pdfstartview=FitB
	}
\usepackage[capitalize]{cleveref}
\usepackage{pgfplots}
\pgfplotsset{compat=newest}

\newcommand{\dg}{\dagger}
\newcommand{\bs}{\boldsymbol}

\newcommand{\mb}{\boldsymbol}
\newcommand{\Mlo}{M_{\rm lo}}
\newcommand{\Mhi}{M_{\rm hi}}

\definecolor{dgreen}{rgb}{0,0.5,0}
\definecolor{cgreen}{rgb}{0,0.8,0}
\definecolor{dblue}{rgb}{0,0,0.7}
\definecolor{dred}{rgb}{0.8,0,0}

\begin{document}

    \title{\texorpdfstring{Halo EFT for $\boldsymbol{^{31}}$Ne in a spherical formalism}{Halo EFT for Ne-31 in a spherical formalism}}

    \author{Wael Elkamhawy}
    \email{elkamhawy@theorie.ikp.physik.tu-darmstadt.de}
    \affiliation{Technische Universität Darmstadt, Department of Physics, Institut f\"ur Kernphysik, 64289 Darmstadt, Germany}
    \affiliation{Helmholtz Forschungsakademie Hessen f\"ur FAIR (HFHF), GSI Helmholtzzentrum f\"{u}r Schwerionenforschung GmbH, 64291 Darmstadt, Germany}
	
    \author{Hans-Werner Hammer}
    \email{Hans-Werner.Hammer@physik.tu-darmstadt.de}
    \affiliation{Technische Universität Darmstadt, Department of Physics, Institut f\"ur Kernphysik, 64289 Darmstadt, Germany}
    \affiliation{Helmholtz Forschungsakademie Hessen f\"ur FAIR (HFHF), GSI Helmholtzzentrum f\"{u}r Schwerionenforschung GmbH, 64291 Darmstadt, Germany}
    \affiliation{ExtreMe Matter Institute EMMI, GSI Helmholtzzentrum f\"{u}r Schwerionenforschung GmbH, 64291 Darmstadt, Germany}
	
    \date{\today}

\begin{abstract}
  We calculate the electromagnetic properties of the deformed
  one-neutron halo candidate $^{31}$Ne using Halo Effective Field
  Theory (Halo EFT). In this framework, $^{31}$Ne is bound via a resonant
  $P$-wave interaction between the $^{30}$Ne core and the valence neutron.
  We set up a spherical formalism for $^{31}$Ne in order to
  calculate the electromagnetic form factors and the E1-breakup
  strength distribution into the $^{30}$Ne-neutron continuum at leading
  order in Halo EFT. The associated uncertainties are estimated according
  to our power counting.
  In particular, we assume that the deformation of the $^{30}$Ne core enters
  at next-to-leading order. It can be accounted for by including the
  $J^P=2^+$ excited state of $^{30}$Ne
  as an explicit field in the effective Lagrangian.
\end{abstract}

\maketitle

\section{Introduction}
The emergence of halo nuclei is an intriguing aspect of atomic
nuclei near the driplines~\cite{Hansen:1995pu,Jonson:2004,Riisager:2012it}.
They were discovered in the 1980s at radioactive beam
facilities and are characterized by an unusually large interaction
radius~\cite{Tanihata:2016zgp}.
Nuclear halo states consist of a tightly bound core with a
characteristic size $\sim 1/\Mhi$ and a cloud of halo nucleons of size
$\sim 1/\Mlo$, which is much larger than neighbouring isotopes.
The large separation between the momentum scales $\Mhi\gg \Mlo$
leads to universal properties, which are independent
of the details of the core
\cite{Jensen:2004zz,Braaten:2004rn,Hammer:2017tjm}.
These properties are most pronounced in neutron
halos as they are not affected by the long-range Coulomb repulsion between
charged particles.

The separation of scales in halo nuclei can formally be exploited using
Halo effective field theory (Halo EFT)~\cite{Bertulani:2002sz,Bedaque:2003wa}
(see Ref.~\cite{Hammer:2017tjm} for a recent review).
It uses effective degrees of freedom and allows to describe 
observables in a systematic expansion in $\Mlo/\Mhi$, thus
enabling uncertainty estimates based on the expected size of
higher order terms in the expansion.
For the dynamics of the halo nucleons, the substructure of the core can
be considered short-distance physics that is not resolved, although low-lying
excited states of the core sometimes have to be included explicitly.
One assumes the core to be structureless and
treats the nucleus as a few-body system of the core and the valence
nucleons.
Corrections from the core structure appear at higher orders in the EFT
expansion, and can be accounted for in perturbation theory.
Since the relevant halo scale $\Mlo$ is small compared
to the pion mass,
even the pion exchange interaction between nucleons
and/or the nuclear core is not resolved.
Thus, halos can be described by an EFT with
short-range contact interactions. 
A new facet compared to few-nucleon systems is the appearance of
resonant interactions in higher partial
waves~\cite{Bertulani:2002sz,Bedaque:2003wa}.
However, there are many light halo nuclei where $S$-wave interactions are
dominant.

In heavier halo nuclei in the region $Z= 9$ -- $12$, the structure of the
ground state is believed to be more complicated and deformed halos are
expected.
Nakamura and collaborators provided the first indications
of a halo structure in $^{31}$Ne~\cite{PhysRevLett.103.262501}.
Subsequently, they showed that the ground state of $^{31}$Ne
has a low one-neutron separation energy and is a deformed $P$-wave
halo~\cite{Nakamura:2014hxa}. A similar structure was found for 
$^{37}$Mg~\cite{Kobayashi:2014owa} which is even heavier.
In the case of $^{31}$Ne,
they used state of the art shell model calculations to analyze nuclear
and electromagnetic $1n$-removal reactions on C and Pb targets and found
that the weakly-bound $P$-wave neutron carries only
about 30\% of the single-particle strength.
The first excitation energy of the $^{30}$Ne core is $800$ keV above the
ground state which has spin and parity quantum numbers $J^P=0^+$.
Meanwhile, the quantum numbers of $^{31}$Ne are $J^P=3/2^-$ with a neutron
separation energy of $150$ keV.

The possibility that $^{31}$Ne could a be a one-neutron halo was
suggested in theoretical work using density-dependent
relativistic mean-field theory~\cite{Zhong_Zhou_2001}. Various authors
have analyzed the experimental data on Coulomb dissociation
of $^{31}$Ne based on this assumption
\cite{PhysRevC.81.021304,PhysRevC.83.041303}.
Urata et al. showed that the data can be well
reproduced in the particle-rotor model
when the quadrupole deformation parameter of the $^{30}$Ne
core is around $\beta_2=0.2...0.3$~\cite{PhysRevC.83.041303}.
The preferred structure of a $P$-wave neutron halo with $J^P=3/2^-$
was also obtained using a microscopic $G$-Matrix calculation
\cite{PhysRevLett.108.052503} and a deformed Woods–Saxon potential for the
neutron-core interaction \cite{SHUBHCHINTAK201499}.
A theoretical analysis of the ground state quantum numbers
$^{31}$Ne based on experimental data on Coulomb breakup and neutron removal
reached the same conclusion~\cite{Hong:2017vul}.
Most recently, the Gamow shell model was applied to
the Neon isotopes $^{26}$Ne$-^{31}$Ne~\cite{Li:2022qzo}. This study
confirmed the $P$-wave neutron halo character of $^{31}$Ne and
suggested that $^{29}$Ne could also be a neutron halo.

The separation of scales in $^{31}$Ne allows for a 
controlled, systematic description of its properties using Halo EFT. 
A first Halo EFT calculation of the electric properties of $^{31}$Ne
based on the general framework of \cite{Hammer:2011ye} was carried out
in~\cite{FBS60.16}.
In this paper, we present a complete discussion of the electromagnetic
structure of  $^{31}$Ne, as well as its E1 breakup.
We describe $^{31}$Ne as a $P$-wave $^{30}$Ne-neutron bound state.
The deformation of the $^{30}$Ne core
enters at next-to-leading order and can be
calculated by including the
$J^P=2^+$ excited state as an explicit field in the effective Lagrangian.
Instead of using the standard Cartesian formulation of the field theory
applied in \cite{FBS60.16}, we introduce a spherical
basis that is ideally suited for the description of halo nuclei beyond
the $S$-wave. It employs the correct number of field components in a
given partial wave and thus does not require any auxilliary conditions,
leading to more compact and transparent expressions. Moreover, we
also calculate magnetic observables.

In Sec.~\ref{sec:HaloEFT}, we present the Halo EFT for $^{31}$Ne in
the spherical basis, derive the $^{30}$Ne-neutron scattering amplitude,
and discuss the corresponding power counting.
The electromagnetic (EM) sector is discussed in Sec.~\ref{sec:EMSector}.
We incorporate EM interactions and derive the scalar and vector currents
and their corresponding form factors. In Sec.~\ref{sec:Correlation},
we extract the leading moments from the form factors and discuss
universal correlations between them. Moreover, we elucidate the
implications of the multipole moments with respect to the deformation
of $^{31}$Ne
and determine the quadrupolar deformation parameter $\beta_2$. The
E$1$ breakup of $^{31}$Ne into the $^{30}$Ne-neutron
continuum is analyzed in Sec.~\ref{sec:E1}. Finally,
we present our conclusions in Sec.~\ref{sec:Conclusion}.

\section{Halo EFT for Neon-31\label{sec:HaloEFT}}

\subsection{Lagrangian: strong sector}
We describe $^{31}$Ne as a shallow $P$-wave bound state of the $^{30}$Ne core
and the valence neutron. Our effective Lagrangian includes a bosonic field 
$c$ with $J^P=0^+$ for the $^{30}$Ne core and a $J^P=1/2^+$ spinor field
$n_{\alpha}$ with $\alpha \in \{-1/2,1/2\}$ for the neutron. Moreover,
a  $J^P=3/2^-$ dimer field $\pi_{\beta}$ with $\beta \in \{-3/2, -1/2, 1/2, 3/2\}$
captures the physics of $^{31}$Ne and the core-neutron continuum.
The corresponding Lagrangian is given by
\begin{equation}
		\begin{aligned}
		\label{eq:Lagrangian}
			\mathcal{L}	= &~c^\dg\left[i\partial_0+\frac{{\nabla}^2}{2m_c} \right]c+n_{\alpha}^\dg \left[ i\partial_0+ 
										\frac{{\nabla}^2}{2m_n} \right] n_{\alpha} + \pi_{\beta}^\dg \left[ \eta_1\left(i\partial_0+\frac{{\nabla}^2}{2M_{nc}}\right)
										+\Delta_1 \right]\pi_{\beta}\\
										&-g_1\left[\left(c^\dg\overleftrightarrow{\nabla}_{\hspace{-0.13cm}i}~n_{\alpha}^\dg\right)\pi_{\beta}~C_{\scriptscriptstyle{(1i)(\frac{1}{2}\alpha)}}
										^{\scriptscriptstyle{\frac{3}{2}\beta}} + \text{H.c.} \right],
		\end{aligned}
	\end{equation}
where $M_{nc}\equiv m_n+m_c$ denotes the kinetic mass of the
core-neutron system, while $\eta _1\equiv \pm 1$ is a sign to be
determined from matching to scattering observables.
Moreover, $\overleftrightarrow{\mb{\nabla}}\equiv m_R\left[{m_c}^{-1}\overleftarrow{\mb{\nabla}}-{m_n}^{-1}\overrightarrow{\mb{\nabla}}\right]$
is the Galilean-invariant
derivative where $m_R=m_c m_n/(m_c+m_n)$ denotes the core-neutron
reduced mass. The coefficient
$C_{\scriptscriptstyle{(1i)(\frac{1}{2}\alpha)}}^{\scriptscriptstyle{\frac{3}{2}\beta}}$ 
is a Clebsch-Gordan coefficient coupling the neutron spin and the
core-neutron
relative angular momentum to the total spin $J=3/2$ of the dimer field.
Note that the index $i$ 
of the derivative operator is a spherical index denoting the projection of
the $P$-wave interaction whereas $\alpha$ and $\beta$ are spinor indices
denoting the projections
of their corresponding spins. 
Moreover, we use spherical coordinates throughout this work. Our
conventions are summarized in Appendix~\ref{sec:AppSpherical}.

\subsection{Full Dimer Propagator}
\label{sec:Dimer}

\begin{figure}[t]
\centering
\includegraphics[scale=0.8]{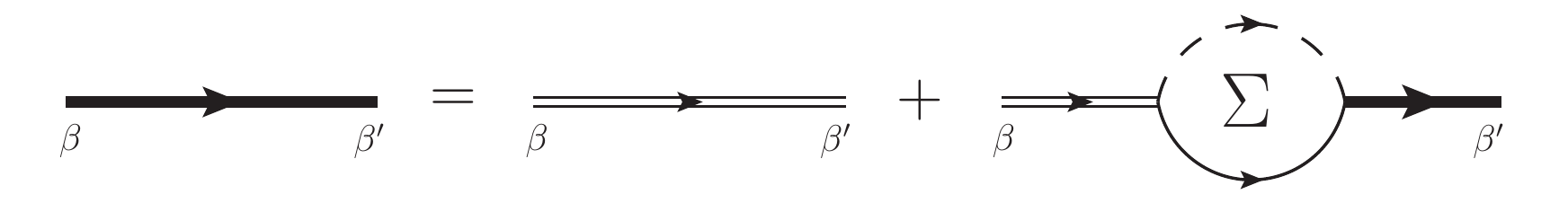}  
\caption{
  Dyson equation for the full dimer propagator.
  The thick line denotes the full dimer propagator, while the double solid
  line denotes the bare propagator. The single solid line represents the
  neutron field, whereas the dashed line represents the core field.}
\label{fig:FullDimeronPropagator}
\end{figure}

For convenience, we use the power divergence subtraction scheme
from Refs.~\cite{Kaplan:1998tg} and~\cite{Kaplan:1998we} with
renormalization scale $\mu$.
In order to determine the full dimer propagator, we dress the bare propagator
\begin{align}
iD^0(p_0,\mb{p})=\frac{i}{\eta_1\left(p_0-\frac{\mb{p}^2}{2M_{nc}}\right)+\Delta_1+i\epsilon}
\end{align}
with dimer self-energies. We end up with the Dyson equation which is depicted
diagramatically in \cref{fig:FullDimeronPropagator}.
This geometric series represents the exact solution of the core-neutron
problem. 
%%%%%%%%%%%%%%%%%%%%%%%%%%%%%%%%%%%%%%%%%%%%%%%%%%%%%%%%%%%%%%%%%%%%%%%%%%%%%
%%%%%%%%%%%%%%%%%%%%%%%%%%%%%%%%%%%%%%%%%%%%%%%%%%%%%%%%%%%%%%%%%%%%%%%%%%

The dimer self-energy is diagonal in the spin indices of the incoming and
outgoing dimer fields and reads
\begin{align}
-i\Sigma_{\beta^{\prime}\beta}(p_0,\mb{p})=-i\Sigma(p_0,\mb{p}) \delta_{\beta^{\prime}\beta},
\end{align}
with
\begin{align}
-i\Sigma(p_0,\mb{p})=\frac{im_R g_1^2 2m_R\left(p_0-\frac{\mb{p}^2}{2M_{nc}}\right)}{6\pi}\left(\frac{3}{2}\mu-\sqrt{-2m_R\left(p_0-\frac{\mb{p}^2}{2M_{nc}}\right)-i\epsilon}\right)\,.
\end{align}
Since the $\Sigma_{\beta^{\prime}\beta}$ and $D^0(p_0,\mb{p})$ are diagonal in
the spin indices, the full dimer propagator is also diagonal and reads	
\begin{equation}
  \begin{aligned}
    iD_{\beta^{\prime}\beta}(p_0,\mb{p})=iD(p_0,\mb{p})\delta_{\beta^{\prime}\beta},
  \end{aligned}
\end{equation}
where the scalar full propagator is given by
\begin{equation}
  \begin{aligned}
    iD(p_0,\mb{p})	&=\frac{iD^0(p_0,\mb{p})}{1-\Sigma(p_0,\mb{p}) D^0(p_0,\mb{p})}\\
    &=\frac{i}{\eta_1\left(p_0-\frac{\mb{p}^2}{2M_{nc}}\right)+\Delta_1-\Sigma(p_0,\mb{p})+i\epsilon}.
    \label{eq:dimerfull}
  \end{aligned}
\end{equation}

The full dimer propagator must have a simple pole at the energy
$p_0=\frac{p^{2}}{2M_{nc}}-B_1$ with $B_1=\gamma_1^{2}/(2m_R)$ the
one-neutron separation energy of $^{31}$Ne, whereas $\gamma_1 >0$ is the
corresponding binding momentum. In order to calculate $^{31}$Ne observables,
we need the wave function renormalization constant defined by
\begin{align}
  Z_\pi^{-1}=\left(\frac{\partial}{\partial p_0}\frac{1}{D(p_0,\mb{p})}\right)\Bigg|_{p_0=\frac{\mb{p}^{2}}{2M_{nc}}-B_1}\,,
\end{align}
which yields
\begin{align}
  Z_\pi=-\frac{6\pi}{m_R^{2}g_1^{2}}\frac{1}{3\gamma_1-\frac{6\pi\eta_1}{m_R^2g_1^2}-3\mu}.
	\end{align}

\subsection{Scattering Amplitude and Matching}

%%%%%%%%%%%%%%%%%%%%%%%%%%%%%%%%%%%%%%%%%%%%%%%%%%%%%%%%%%%%%%%%%%%%%%%%%%
\begin{figure}[t]
  \centering
  \includegraphics[scale=0.5]{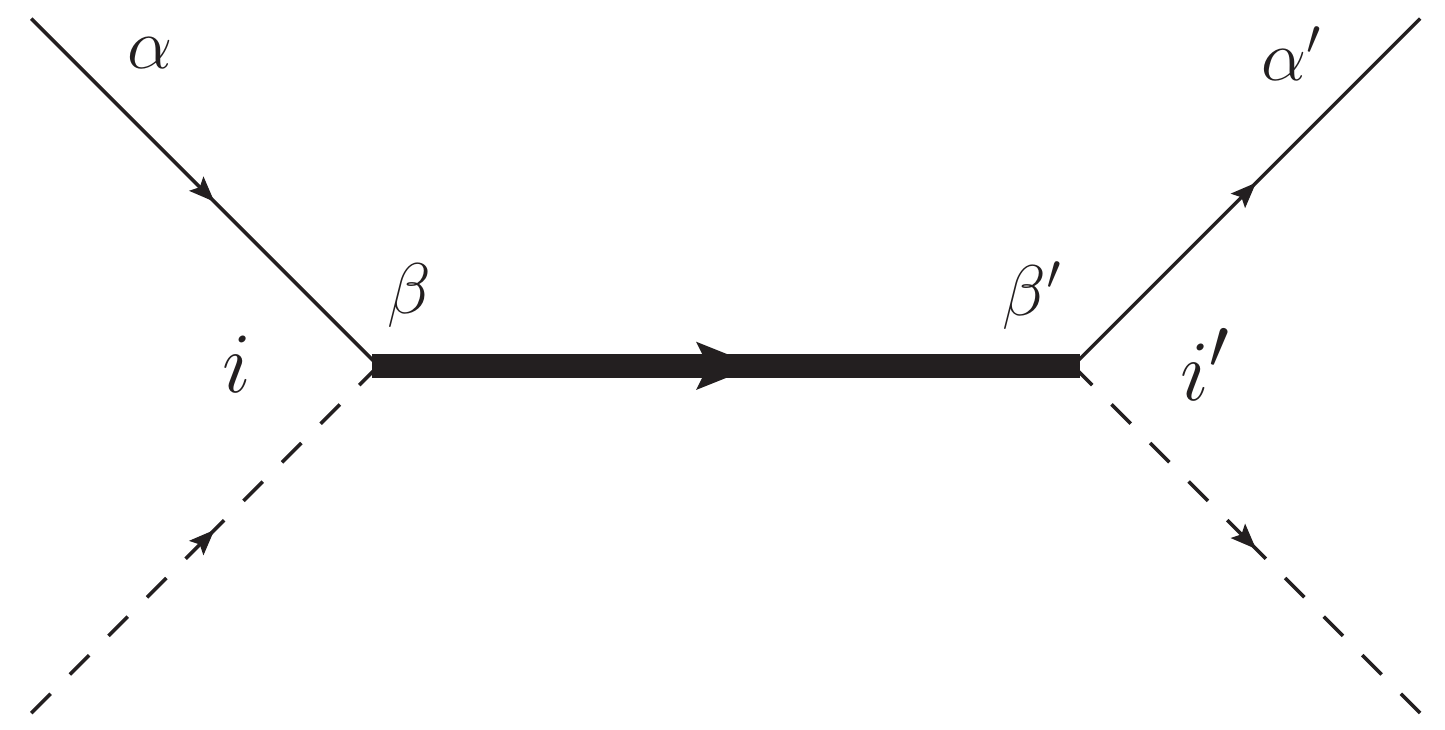}  
  \caption{\label{fig:SA31Ne}Pictoral representation of the neutron-core
    scattering amplitude including spin indices.
    The thick line denotes the full dimer propagator, the single solid line
    represents the neutron field, and the dashed line represents the core field.
  }
  \label{fig:ScatteringAmplitude}
\end{figure}
%%%%%%%%%%%%%%%%%%%%%%%%%%%%%%%%%%%%%%%%%%%%%%%%%%%%%%%%%%%%%%%%%%%%%%%%%

The $P$-wave neutron-core scattering amplitude in the $J=3/2$-channel
is obtained by attaching external
core and neutron lines to the full dimer propagator from
Eq.~\eqref{eq:dimerfull}, see \cref{fig:ScatteringAmplitude}.
In the center-of-mass frame with $E=p^2/(2m_R)=p'^2/(2m_R)$ and
$p=|\mb{p}|=|\mb{p}'|$, it  reads
\begin{align}
  \label{eq:J32eft}
  T_{\alpha^{\prime}\alpha}(\mb{p}',\mb{p})	=\frac{6\pi}{m_R}~\frac{\frac{2}{3}\,\mb{p}'\cdot\mb{p}\,\delta_{\alpha'\alpha}-\frac{i}{3}
    \Big(\mb{\sigma} \cdot \left(\mb{p}'\times\mb{p}\right)\Big)_{\hspace{-0.1cm}\alpha'\alpha}}
  {\left(\frac{6\pi\Delta_1}{m_R g_1^2}+\frac{3\pi\eta_1}{m_R^2 g_1^2}p^2+\frac{3}{2}\mu p^2+ip^3\right)},
\end{align}
where $\mb{\sigma}$ is the three-dimensional vector with the Pauli matrices
as its components. 
The corresponding result for a $J=1/2$ state is given in
\mbox{Appendix \ref{sec:AppJ12}}.

Comparing Eq.~\eqref{eq:J32eft} to the general form of the amplitude
in terms of the  effective range parameters,
	\begin{align}
	  T_{\alpha^{\prime}\alpha}(\mb{p}',\mb{p})	&=\frac{6\pi}{m_R}~\frac{\frac{2}{3}\,\mb{p}'\cdot\mb{p}\,\delta_{\alpha'\alpha}-\frac{i}{3}
	    \Big(\mb{\sigma} \cdot \left(\mb{p}'\times\mb{p}\right)\Big)_{\hspace{-0.1cm}\alpha'\alpha}}
	  {\left(\frac{1}{a_1}-\frac{1}{2}r_1p^2+ip^3\right)}\,,
	\end{align}
we obtain the matching conditions
\begin{align}
  a_1	&=\frac{m_R g_1^2}{6\pi\Delta_1}, \quad%\\[0.5em]
  r_1	=-\frac{6\pi\eta_1}{m_R^2 g_1^2}-3\mu\,.
  \label{eq:ERP}
\end{align}
Since  the parameter $r_1$ has to be negative for causal
scattering~\cite{Hammer:2009zh,Hammer:2010fw},
the sign $\eta_1$ is determined to be $\eta_1 = +1$ according to \cref{eq:ERP}.
With these matching conditions, the wave function renormalization
constant reads
\begin{align}
  \label{eq:Zfull}
  Z_\pi=-\frac{6\pi}{m_R^{2}g_1^{2}}\frac{1}{3\gamma_1+r_1}\,.
\end{align}
$Z_\pi$ is the residue of the bound state pole at the energy
\mbox{$p_0=p^{2}/(2M_{nc})-\gamma_1^{2}/(2m_R)$} in the full dimer
propagator where $\gamma_1$ is a real positive solution to
the equation
\begin{align}
  \frac{1}{a_1}+\frac{r_1}{2}\gamma_1^2+\gamma_1^3=0\,.
  \label{eq:polefull}
\end{align}

\subsection{Power Counting}
Following Ref.~\cite{Bedaque:2003wa}, we assume only one combination of coupling constants to be fine-tuned, namely $\Delta_1/g_1^{2}$. This is sufficient in order to 
produce a shallow $P$-wave bound state. With this choice, the scattering volume $a_1$ is enhanced by
	\begin{align}
		a_1\sim\frac{1}{M_{\text{lo}}^{2}M_{\text{hi}}},
	\end{align}
whereas the $P$-wave effective momentum $r_1$ scales like
	\begin{align}
		- r_1\sim M_{\text{hi}},
		\label{eq:r1-scaling}
	\end{align}
where $M_{\text{lo}}$ and $M_{\text{hi}}$ denote the typical low- and high-momentum scales of the system, respectively. The low-momentum scale is given by
the binding momentum of the shallow $P$-wave bound state. In the case of $^{31}$Ne with a binding energy of $B_1=150$ keV~\cite{Nakamura:2014hxa}, this yields
	\begin{align}
		M_{\text{lo}}\approx\gamma_1=\sqrt{2m_R B_1}=16.5 \text{ MeV}.
		\label{eq:Mlow}
	\end{align}
The high-momentum scale can be approximated by the breakdown scale of the theory. Since we do not include the $J^P=2^+$ state of the $^{30}$Ne core explicitly, $M_{\text{hi}}$ can be estimated by the associated
momentum scale of the excitation energy $E_{\text{ex}}=792$ keV~\cite{ShamsuzzohaBasunia:2010ggy}. 
The corresponding value is given by $M_{\text{hi}}\approx\sqrt{2m_R E_{\text{ex}}}\approx40 \text{ MeV}$. At this momentum scale, the deformation of the $^{30}$Ne core due to this excited state starts to play a role.

Given this power counting scheme, the equation for the pole position,
Eq.~\eqref{eq:polefull}, and the wave function renormalization,
Eq.~\eqref{eq:Zfull}, can be expanded at leading order in $\Mlo/\Mhi$
to yield
\begin{align}
  \gamma_1=\sqrt{-\frac{2}{a_1 r_1}}
\quad \mbox{ and } \quad
  Z_\pi^{\text{LO}}=-\frac{6\pi}{m_R^{2}g_1^{2}r_1}.
		\label{eq:LOresidue}
	\end{align}
        The quantity $g_1^{2}Z_\pi^{\text{LO}}$ is proportional to the absolute value squared of the EFT wave function at the bound state pole.
        Thus, $Z_\pi$ must be positive. % to ensure a normalizable state.
        As a consequence, $r_1$ must be negative. At NLO, the wave function renormalization is given by the expression from Eq.~\eqref{eq:Zfull},
	\begin{align}
		Z_\pi^{\text{NLO}}=-\frac{6\pi}{m_R^{2}g_1^{2}}\frac{1}{3\gamma_1+r_1}.
		\label{eq:NLOresidue}
	\end{align}
        Thus, only if $|r_1|>3\gamma_1$ holds, we end up with a normalizable state with positive residue.
        This requirement is consistent with the hierarchy $\Mlo \ll\Mhi$ which forms the basis of our power counting. 

For a shallow $P$-wave state, we have at least two effective range expansion parameters, $a_1$ and $r_1$, which have to be fixed by observables.
Until now, we only know the neutron separation energy of $^{31}$Ne from experiment, which is not enough in order to fix both effective range expansion parameters. Therefore, we will 
estimate the $P$-wave effective momentum $r_1$ in an interval around the breakdown scale, according to \cref{eq:r1-scaling} and use the neutron separation energy 
to determine the scattering volume $a_1$. Based on these assumptions we can calculate other observables accessible in our theory, such as the electromagnetic current 
and the corresponding multipole moments as well as the associated radii.

Taking everything together, we estimate
\begin{align}
  r_1\sim M_{\text{hi}}\in\left[-150,-50\right]\, \text{MeV}\,.
  \label{eq:r1-estimation}
\end{align}
Values of $r_1$ in this interval are consistent with unitarity and the estimated breakdown scale of our theory
associated with the $J^P=2^+$ excited state of the core. 

\section{Electromagnetic Sector\label{sec:EMSector}}
We now go on to include electromagnetic interactions in the effective theory. Moreover, we derive the corresponding form factors using spherical coordinates. We present results for the form factors of $^{31}$Ne and provide general expressions
for form factors of arbitrary multipolarity $L$. 

In the first step, electromagnetic interactions are included via minimal substitution
	\begin{align}
		\partial_{\mu}\rightarrow D_{\mu}=\partial_{\mu}+ie\hat{q}A_{\mu},
	\end{align}
meaning that the usual derivative $\partial_{\mu}$ in the Lagrangian in \cref{eq:Lagrangian} is replaced by the covariant derivative $D_{\mu}$ containing the charge 
operator $\hat{q}$, the elementary charge $e>0$ and the photon field $A^{\mu}=(A_0,\mb{A})$. In the second step, all possible gauge-invariant operators involving the electric 
field $\mb{E}$ and also the magnetic field $\mb{B}$ have to be considered within our power counting scheme. It turns out that only gauge-invariant operators 
proportional to the magnetic field $\mb{B}$ are contributing at LO whereas operators involving the electric field $\mb{E}$ contribute at higher orders.
\subsection{Scalar Current}
\begin{figure}[t]
		\centering
		\includegraphics[scale=1]{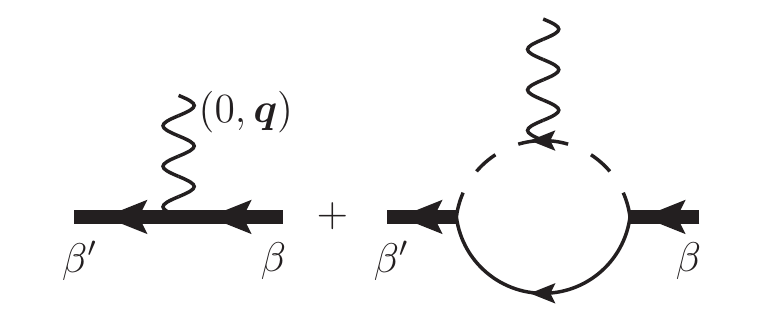}  
		\caption{Diagrams contributing to the irreducible vertex for an $A_0$ photon coupling to the $^{30}$Ne-$n$ $P$-wave bound state at LO.}
		\label{fig:ElectricFormFactor}
	\end{figure}

First, we calculate the matrix element of the zeroth component of the electromagnetic current of $^{31}$Ne. Therefore, we consider the amplitude with an irreducible 
vertex for an $A_0$ photon with four momentum $(0,\mb{q})$ coupling to the $^{30}$Ne-$n$ $P$-wave bound state
with initial momentum $\mb{p}$ and final momentum $\mb{p}'$. Thus, we have $\mb{q}=\mb{p}'-\mb{p}$ and
define $q=|\mb{q}|$. 
The initial and final states are characterized by their momenta and projections of the spin, denoted by $\ket{\pi_{\beta}(\mb{p})}$ and $\ket{\pi_{\beta'}(\mb{p}')}$, 
respectively. The LO contributions to this amplitude are depicted in \cref{fig:ElectricFormFactor}.
Since $^{31}$Ne has a total spin of $3/2$, there are four possible projections for each the initial and final state. Hence, the tensors connecting initial and final
state projections are $4\times4$ matrices in spin space. 
%%%%%%%%%%%%%%%%%%%%%%%%%%%%%%%%%%%%%%%%%%%%%%%%%%%%%%%%%%%%%%%%%%
%%%%%%%%%%%%%%%%%%%%%%%%%%%%%%%%%%%%%%%%%%%%%%%%%%%%%%%%%%%%%%%%%%%%%%

        The scalar electromagnetic transition amplitude 
        can be written as
	\begin{equation}
		\begin{aligned}
			\label{eq:ScalarFormFactor}
			\bra{\pi_{\beta'}(\mb{p}')}J^{0}\ket{\pi_{\beta}(\mb{p})}=&-iq_c eG_{\text{E}0}(q)\sqrt{\frac{4\pi}{1}}q^{0}Y_{00}^*(\mb{e}_{\mb{q}})
			\left(\tilde{T}^{\left[00\right]}_{3/2}\right)_{\beta'\beta}\\
			&-i\mu_QG_{\text{E}2}(q)\frac{1}{2}\sqrt{\frac{4\pi}{5}}q^{2} \sum_M Y_{2M}^*(\mb{e}_{\mb{q}})\left(\tilde{T}^{\left[2M\right]}_{3/2}\right)_{\beta'\beta},
		\end{aligned}
	\end{equation}
where $\bs{e}_{\bs{q}}$ is the unit vector of $\bs{q}$, $q_c$ is the charge of the core in terms of the elementary charge $e$ and $\mu_Q$ is the quadrupole moment. Moreover, the electric monopole and quadrupole form factors 
are denoted by $G_{\text{E}0}(q)$ and $G_{\text{E}2}(q)$, respectively. Since the multipole moments are explicitly factored out of the form factors in \cref{eq:ScalarFormFactor},  $G_{\text{E}0}(q)$ and $G_{\text{E}2}(q)$ are normalized to one in the limit of vanishing photon momentum by construction. The tensors $\tilde{T}^{00}_{3/2}$ and $\tilde{T}^{2M}_{3/2}$ are normalized $4\times4$ 
polarization matrices~\cite{Khersonskii:1988krb}.
For general $J$, they are $(2J+1)\times(2J+1)$ matrices given by
	\begin{align}
		\left(\tilde{T}^{\left[LM\right]}_{J}\right)_{\beta'\beta}=\frac{C_{\scriptscriptstyle{(J\beta)(LM)}}^{\scriptscriptstyle{J\beta'}}}
		{C_{\scriptscriptstyle{(JJ)(L0)}}^{\scriptscriptstyle{JJ}}}.
	\end{align}
They are normalized such that they have a coefficient of $1$ for maximal projections. Consequently, the multipole moments are defined for maximal projections as it 
is usually done by convention. The subscript $J$ indicates the spin of the considered two-particle bound state, while $L$ stands for the angular momentum of the photon. The possible contributions for $L$
result from coupling the two $P$-wave spherical harmonics appearing in the right diagram of \cref{fig:ElectricFormFactor}. Furthermore, $M$ denotes the 
projection of the angular momentum $L$. This  implicit angular momentum
coupling can yield contributions of the photon multipolarities $L=0,1,2$
in the electromagnetic transition amplitude, Eq.~\eqref{eq:ScalarFormFactor}.
        Due to parity conservation, however, only even numbers of $L$ contribute so that we are left with $L\in\{0,2\}$.
        Hence, as we can read off \cref{eq:ScalarFormFactor}, the electric monopole and quadrupole form factors
        with their corresponding multipole moments appear for $J=3/2$,
        but no dipole form factor.\footnote{Note that if we would consider a spin $1/2$ dimer,
        there would be no quadrupole contribution in \cref{eq:ScalarFormFactor} because of the properties of the Clebsch-Gordan coefficients.}
The LO results for the electric form factors read
	\begin{align}
		\label{eq:MonopoleFormFactor}
		G_{\text{E}0}(q)&=\left[1-\frac{\gamma_1}{r_1}+\frac{y^{2}q^{2}+2\gamma_1^{2}}{yqr_1}\arctan\left(\frac{yq}{2\gamma_1}\right)\right],\\
		\label{eq:QuadrupoleFormFactor}
		\mu_Q G_{\text{E}2}(q)&=-\frac{q_{c}e}{2r_1yq^{3}}\left[2\gamma_1 yq+\left(y^{2}q^{2}-4\gamma_1^{2}\right)\arctan\left(\frac{yq}{2\gamma_1}\right)\right],
	\end{align}
        with
\begin{equation}
  y=m_n/M_{nc}=m_R/m_c\,.
  \label{eq:defy}
\end{equation}
Since gauge invariance ensures charge conservation, the normalization
$\lim\limits_{q \to 0}G_{\text{E}0}(q)= 1$ is automatically
fulfilled and hence serves as a consistency check.
The normalization condition $\lim\limits_{q \to 0}G_{\text{E}2}(q)=1$
determines the quadrupole moment
	\begin{align}
		\label{eq:QuadrupoleMoment}
		\mu_Q=-\frac{y^{2}q_{c}e}{3\gamma_1 r_1}.
	\end{align}
        Inserting this quadrupole moment in \cref{eq:QuadrupoleFormFactor}, we
        obtain
	\begin{align}
		\label{eq:NormalizedQuadrupoleFormFactor}
		G_{\text{E}2}(q)&=\frac{3\gamma_1}{2y^{3}q^{3}}\left[2\gamma_1 yq+\left(y^{2}q^{2}-4\gamma_1^{2}\right) \arctan\left(\frac{yq}{2\gamma_1}\right)\right].
	\end{align}
        Note that the result for $G_{\text{E}0}(q)$, \cref{eq:MonopoleFormFactor}, is the same for a spin $1/2$ dimer but appears with
        the appropriate $J=3/2$  polarization matrix in \cref{eq:ScalarFormFactor}.   
\subsection{Vector Current}
%%%%%%%%%%%%%%%%%%%%%%%%%%%%%%%%%%%%%%%%%%%%%%%%%%%%%%%%%%%%%%%%%%%%%%%%
\begin{figure}[t]
		\centering
		\includegraphics[scale=0.9]{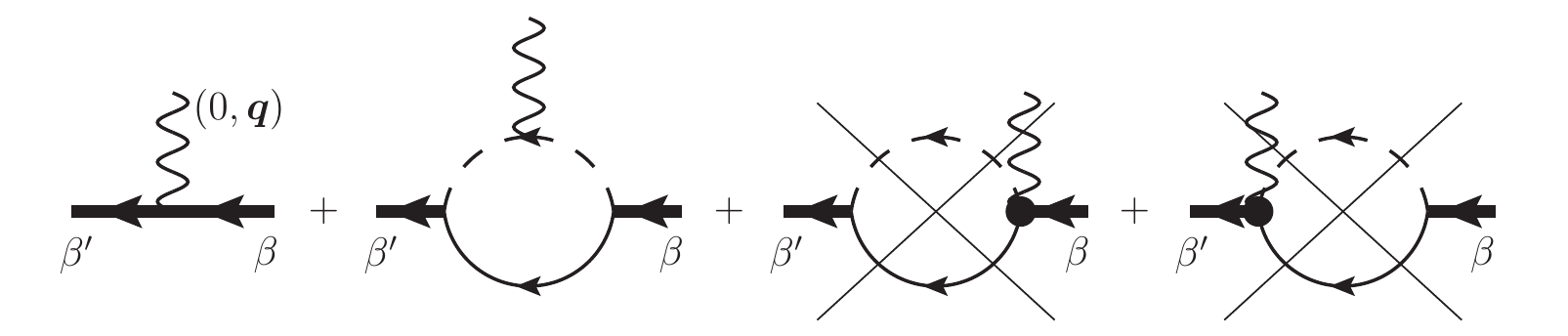}  
		\caption{Diagrams contributing to the irreducible vertex for an $A_k$ photon coupling to the $^{30}$Ne-$n$ $P$-wave bound state at LO. The two
                  diagrams on the right, which are crossed out, can be shown to vanish by parity conservation.}
		\label{fig:VectorPhotons}
\end{figure}
\begin{figure}[t]
		\centering
		\includegraphics[scale=1]{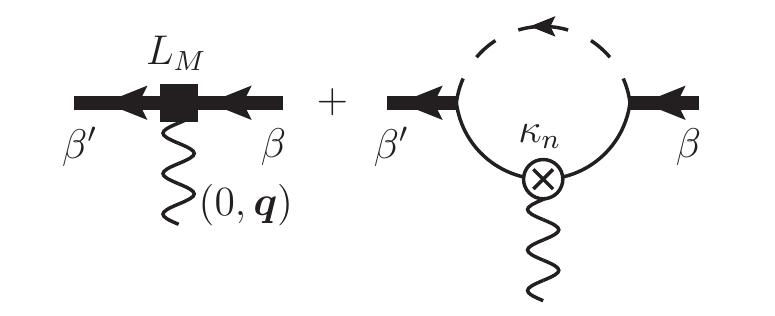}  
		\caption{Diagrams contributing to the irreducible vertex for a $B_k$ photon coupling to the $^{30}$Ne-$n$ $P$-wave bound state at LO. On the one hand, the magnetic
		photon can couple to the neutron spin via the anomalous magnetic moment $\kappa_n$ and on the other hand it can directly couple to the dimer spin via the counterterm
		$L_{\text{M}}$.}
		\label{fig:MagneticPhotons}
\end{figure}

%%%%%%%%%%%%%%%%%%%%%%%%%%%%%%%%%%%%%%%%%%%%%%%%%%%%%%%%%%%%%%%%%%%%%%%
        Next, we investigate the vector electromagnetic current of $^{31}$Ne.
        For this purpose, we consider the amplitude with an irreducible vertex for an $A_k$ photon with 
four momentum $(0,\mb{q})$ coupling to the $^{30}$Ne-$n$ $P$-wave bound state. The corresponding diagrams are depicted in \cref{fig:VectorPhotons}. 
The two diagrams on the right, which are crossed out, can be shown to vanish by parity conservation such that only the two diagrams on the left contribute.
Furthermore, we have to take into account local gauge invariant contributions from the magnetic coupling to the spins of the corresponding fields. Assuming that both the anomalous magnetic moment of the neutron $\kappa_n$ and the magnetic moment of the dimer $L_{\text{M}}$ scale naturally, they contribute at LO. As a matter of fact, the counterterm $L_{\text{M}}$ is necessary for renormalization purposes already at LO.
  The diagrams are shown in \cref{fig:MagneticPhotons} and the
corresponding magnetic interaction vertices are given by
	\begin{align}
		\label{eq:MOpn}
		\mathcal{L}_{\text{M}}^{n}&=\kappa_n \mu_N ~n^{\dg}_{\alpha'}\left(2\mb{S}_{\frac{1}{2}}\cdot\mb{B}\right)_{\hspace{-0.15cm}\alpha'\alpha}n_{\alpha},\\
		\label{eq:MOpPi}
		\mathcal{L}_{\text{M}}^{\pi}&=L_{\text{M}}\mu_N ~\pi^{\dg}_{\beta'}\left(\frac{2}{3}\mb{S}_{\frac{3}{2}}\cdot\mb{B}\right)_{\hspace{-0.15cm}\beta'\beta}\pi_{\beta},
	\end{align}
        where $\kappa_n$ denotes the anomalous magnetic moment of the neutron  and $\mu_N$ is the nuclear magneton. (See Ref.~\cite{Fernando:2015jyd} for a discussion of the $J^P=1/2^+$ case.)
$L_{\text{M}}$ is a counterterm required for renormalization. Furthermore, 
$\mb{B}=\left(\mb{\nabla} \times \mb{A}\right)$ is the magnetic field of the photon while $\mb{S}_J$ is a three-dimensional vector with spin matrices as its components. These matrices depend on the total spin $J$ of the considered field as indicated by the subscript $J$. In our case, $J=1/2$ for the neutron while $J=3/2$ for the dimer field. 
The magnetic operators in \cref{eq:MOpn,eq:MOpPi} are written in the form 
	\begin{align}
		\frac{1}{J}\mb{S}_J \cdot \mb{B}=\sum_{m=-1}^{1} \frac{1}{J} \left(S_J\right)_m B^m\,,
	\end{align}
        where $\left(S_J\right)_m$ and $B_m$ are the corresponding components in spherical coordinates.
        The components of the three spin matrices are given by \cite{Khersonskii:1988krb}
	\begin{align}
		\left[\left(S_J)\right)_m\right]_{\sigma'\sigma}=\sqrt{J(J+1)}C_{\scriptscriptstyle{(J\sigma)(1m)}}^{\scriptscriptstyle{J\sigma'}}.
	\end{align}
Therefore, the matrix element for maximal projection is always multiplied by $B_0$. 

%%%%%%%%%%%%%%%%%%%%%%%%%%%%%%%%%%%%%%%%%%%%%%%%%%%%%%%%%%%%%%%%%%%%%%%%%%%
%%%%%%%%%%%%%%%%%%%%%%%%%%%%%%%%%%%%%%%%%%%%%%%%%%%%%%%%%%%%%%%%%%%%%%

The vector electromagnetic transition amplitude 
can be written as
\begin{equation}
  \begin{aligned}
    \label{eq:VectorFormFactor}
    \bra{\pi_{\beta'}(\mb{p}')}J^{k}\ket{\pi_{\beta}(\mb{p})}=&\left[iq_{c}eG_{\text{E}0}(q)\sqrt{\frac{4\pi}{1}}q^{0}Y_{00}^*(\mb{e}_{\mb{q}})\left(\tilde{T}^{\left[00\right]}_{3/2}
    \right)_{\beta'\beta}\right.\\ 
    &\left.\phantom{\left[ \right.}+i \mu_Q G_{\text{E}2}(q)\frac{1}{2}\sqrt{\frac{4\pi}{5}}q^{2}\sum_M Y_{2M}^*(\mb{e}_{\mb{q}})\left(\tilde{T}^{\left[2M\right]}_{3/2}\right)_{\beta'\beta} \right]
    \frac{\left(\mb{p}'+\mb{p}\right)^{k}}{2M_{nc}}\\
    &+i\mu_D G_{\text{M}1}(q)\sqrt{\frac{4\pi}{3}}q^{1}\sum_M \sqrt{2} C_{\scriptscriptstyle{(1k)(1M)}}^{\scriptscriptstyle{1(M+k)}} Y_{1(M+k)}^*(\mb{e}_{\mb{q}}) 
    \left(\tilde{T}^{\left[1M\right]}_{3/2}\right)_{\beta'\beta}\\
    &+i\mu_O G_{\text{M}3}(q)\frac{1}{2}\sqrt{\frac{4\pi}{7}}q^{3}\sum_M \sqrt{2} C_{\scriptscriptstyle{(1k)(3M)}}^{\scriptscriptstyle{3(M+k)}} Y_{3(M+k)}^*
    (\mb{e}_{\mb{q}}) \left(\tilde{T}^{\left[3M\right]}_{3/2}\right)_{\beta'\beta},
  \end{aligned}
\end{equation}
where $\mu_D$ denotes the magnetic dipole moment, $\mu_O$ is the magnetic octupole moment, whereas $G_{\text{M}1}$ and $G_{\text{M}3}$ are the corresponding
form factors, respectively. As for the scalar current, the multipole moments are explicitly factored out in \cref{eq:VectorFormFactor} and therefore the form factors are normalized to one in the limit of vanishing photon momentum by construction. Obviously, the physics of the vector electromagnetic current is richer than that of the scalar current. Not only the electric monopole and quadrupole form factor appear in 
\cref{eq:VectorFormFactor} but also the magnetic contributions. In our case, or rather for a spin-$3/2$ particle,
in addition to the magnetic dipole moment there is also a 
magnetic octupole moment.
Due to the integral over the two $P$-wave spherical harmonics and over the spherical harmonic from the vector photon in the second diagram of 
\cref{fig:VectorPhotons}, we now have an implicit coupling of three angular momenta. The corresponding
photon multipolarities are $L=0,1,2,3$. 
However, parity conservation restricts the possible values for the magnetic contributions to $L\in\{1,3\}$. The contributions of $L=1$
and $L=3$ manifest in \cref{eq:VectorFormFactor} through terms proportional to $Y_{1(M+k)}^*$ and $Y_{3(M+k)}^*$, respectively. In contrast, the electric contributions to the vector 
current are apparent through the term proportional to $\left(\mb{p}'+\mb{p}\right)^k$. 

A closer look at the vector current in \cref{eq:VectorFormFactor} suggests a generalized structure including arbitrary high multipole electric and magnetic form factors that can be found in \mbox{Appendix \ref{sec:AppGeneralized}}.

The LO results for the magnetic form factors for a $J=3/2$ dimer read
	\begin{align}
		\label{eq:DipoleFormFactor}
		\mu_D G_{\text{M}1}(q)&=\left[a L_{\text{M}}(\mu) + b \kappa_n + c \frac{yq_c}{A_c}\right]\frac{\mu_N}{r_1},\\
		\label{eq:OctupoleFormFactor}
		\mu_O G_{\text{M}3}(q)&=\frac{-\sqrt{6}\kappa_n\mu_N}{10r_1(1-y)q^{3}}\left[2\gamma_1 (1-y)q+\left((1-y)^{2}q^{2}-4\gamma_1^{2}\right) \arctan\left(\frac{(1-y)q}{2\gamma_1}
		\right)\right],
	\end{align}
with $A_c$ denoting the mass number of the core while $a,b$ and $c$ are functions given by
	\begin{align}
    \label{eq:a}
		a&=\left[r_1+3\left(\mu-\gamma_1\right)\right],\\
		\label{eq:b}
		b&=\frac{1}{(1-y)q}\left[\frac{21}{10}\gamma_1(1-y)q+\left(\frac{21}{20}(1-y)^{2}q^{2}+\frac{9}{5}\gamma_1^{2}\right)\arctan\left(\frac{(1-y)q}{2\gamma_1}\right)\right]-3\mu,\\
		\label{eq:c}
		c&=\frac{3}{4q}\left[2\gamma_1 q+\frac{\left(4\gamma_1^{2}+y^{2}q^{2}\right)}{y}\arctan\left(\frac{yq}{2\gamma_1}\right)\right]-3\mu\,.
	\end{align}
        For a $J=1/2$ dimer, only the magnetic dipole form factor is observable.
        The corresponding functions $a$, $b$ and $c$ slightly differ from the ones given in \cref{eq:a,eq:b,eq:c} and can be found in \cref{sec:AppJ12}.
The first term in the numerator of \cref{eq:DipoleFormFactor} proportional to $L_{\text{M}}(\mu)$ is a contribution due to the direct magnetic moment coupling to the spin of the dimer field. The second term is a contribution due to the magnetic moment of the neutron proportional to $\kappa_n$. Finally, the origin of the third contribution proportional to $yq_c/A_c$ lies in the finite angular momentum of the charged core which induces a 
magnetic dipole moment and therefore contributes to the magnetic dipole form factor.

Applying the normalization conditions at the real photon point,
$\lim\limits_{q \to 0}G_{\text{M}1}(q)= 1$ and
$\lim\limits_{q \to 0}G_{\text{M}3}(q)= 1$,
the magnetic dipole and octupole moments are read off as
	\begin{align}
		\label{eq:DipoleMoment}
		\mu_D&=\left[L_{\text{M}}(\mu)+\frac{3(\mu-\gamma_1)}{r_1}\left(L_{\text{M}}(\mu)-\kappa_n-\frac{yq_c}{A_c}\right)\right]\mu_{N},\\
		\label{eq:OctupoleMoment}
		\mu_O&=-\frac{\sqrt{6}(1-y)^{2}\kappa_n \mu_N}{15\gamma_1 r_1}.
	\end{align}

Note that the magnetic dipole moment cannot be predicted in Halo EFT. Instead
the counterterm $L_{\text{M}}(\mu)$ is matched to the magnetic dipole moment
using Eq.~(\ref{eq:DipoleMoment}). The contribution of the dimer
to the magnetic moment, $L_{\text{M}}(\mu)$,
thus is resolution dependent. Its scale
dependence is governed by the renormalization group equation
\begin{align}
  \label{eq:LMmu}
  \mu\frac{d}{d\mu}\,L_{\text{M}}(\mu) = \frac{\mu}{\mu-\gamma_1+r_1/3}\left[
      \kappa_n+\frac{yq_c}{A_c} -L_{\text{M}}(\mu)\right]\,.
\end{align}
However, the full $q^2$-dependence of the product
$\mu_D G_{\text{M}1}(q)$ in the transition amplitude, Eq.~(\ref{eq:VectorFormFactor}) is given by Halo EFT. Moreover, the octupole moment, in contrast to
the  dipole moment, is also predicted by Halo EFT.
Inserting \cref{eq:OctupoleMoment} in \cref{eq:OctupoleFormFactor} yields
	\begin{align}
		\label{eq:NormalizedOctupoleFormFactor}
		G_{\text{M}3}(q)&=\frac{3\gamma_1}{2(1-y)^{3}q^{3}}\left[2\gamma_1 (1-y)q+\left((1-y)^{2}q^{2}-4\gamma_1^{2}\right) \arctan\left(\frac{(1-y)q}{2\gamma_1}\right)\right]\\
                \nonumber
		&=G_{\text{E}2}(q)\, [y\rightarrow(1-y)].
	\end{align}
        This is the exact same expression as in \cref{eq:NormalizedQuadrupoleFormFactor} except for the substitution $y\rightarrow(1-y)$
in the dependence on the mass factor from \cref{eq:defy}.

\section{Bound State Observables and Their Correlations\label{sec:Correlation}}

%%%%%%%%%%%%%%%%%%%%%%%%%%%%%%%%%%%%%%%%%%%%%%%%%%%%%%%%%%%%%%%%%%%%%%%
\begin{figure}[t]
    \centering
    \includegraphics[width = 0.49\textwidth]{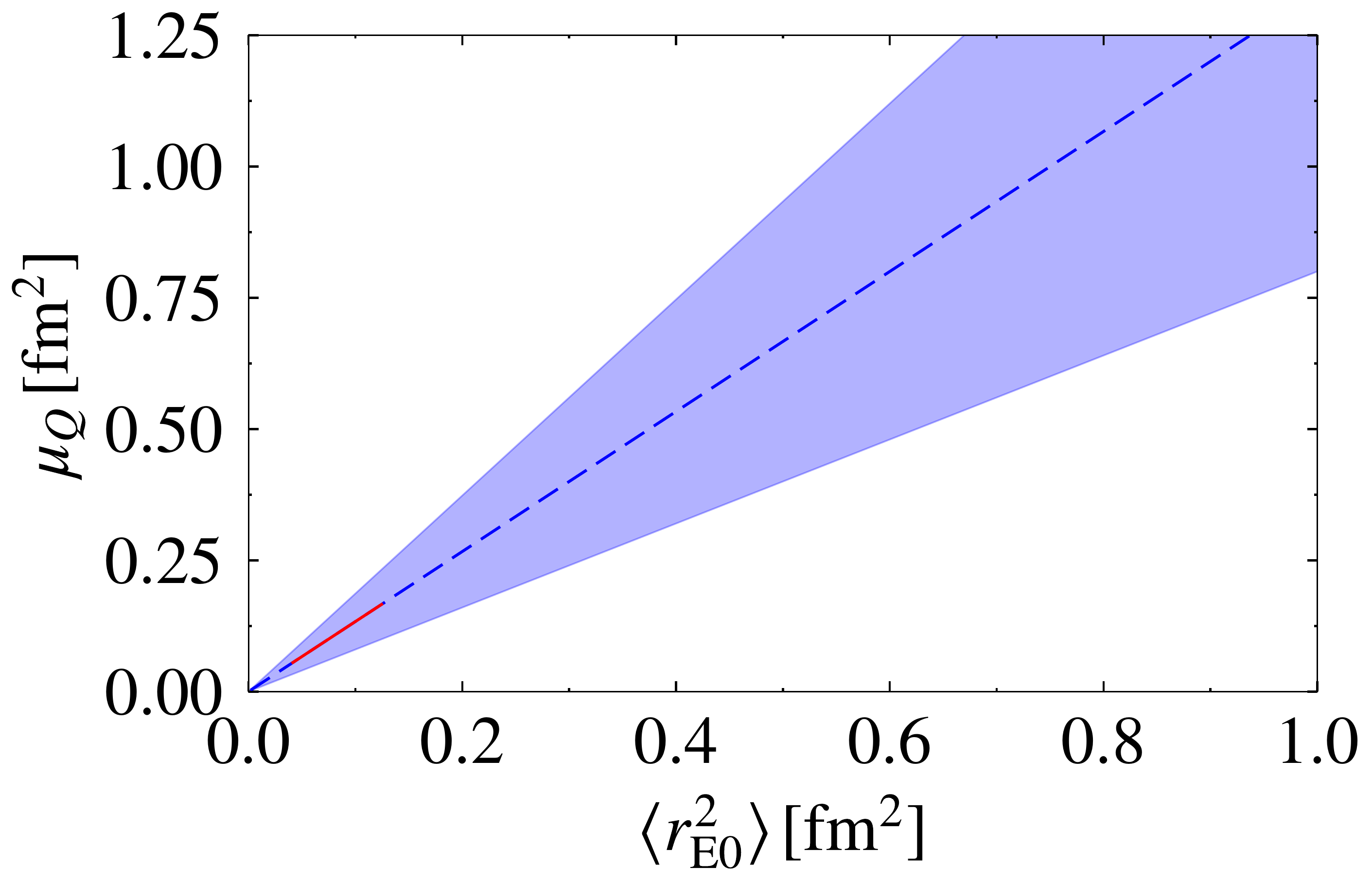}
    \includegraphics[width = 0.49\textwidth]{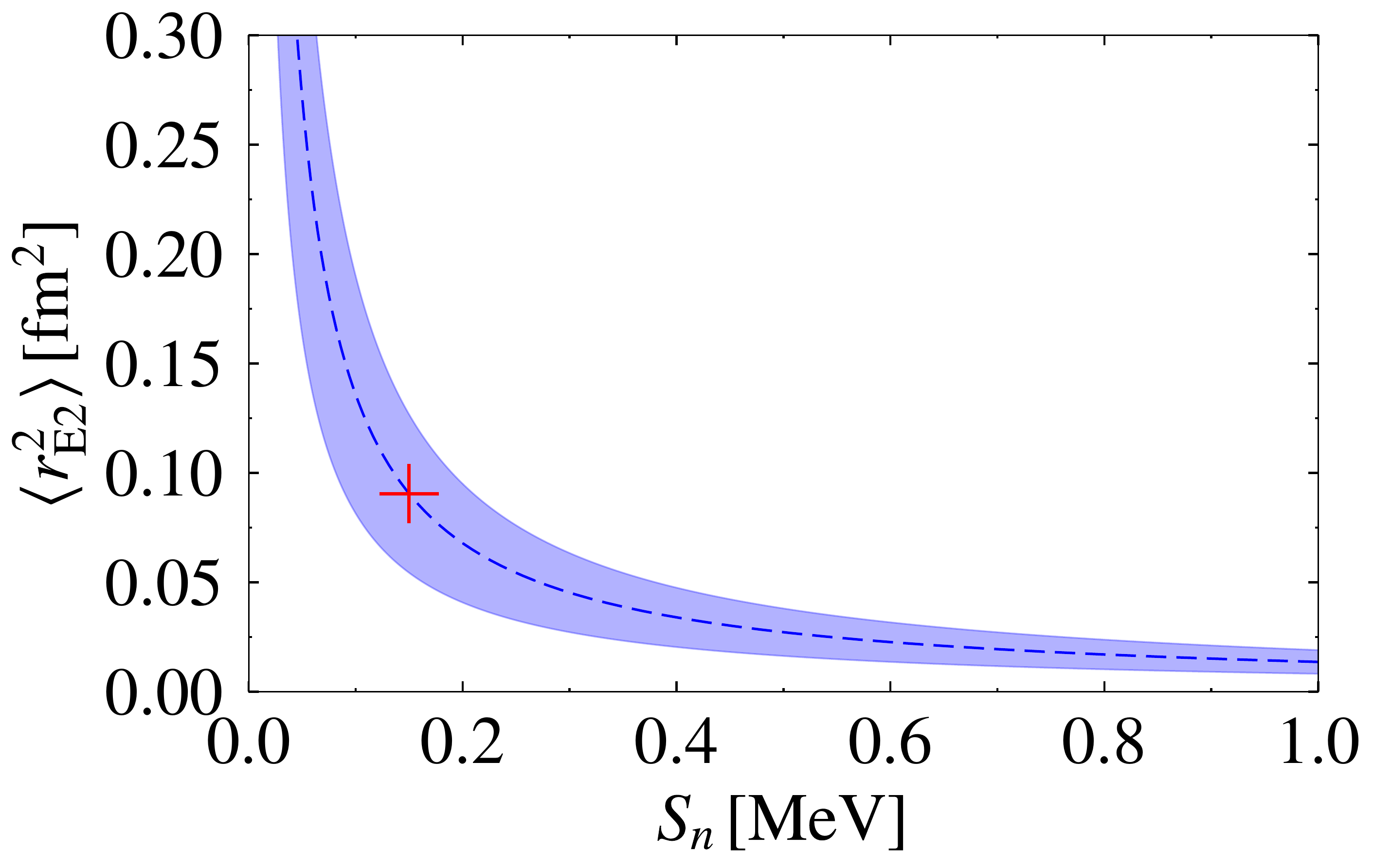}
    \caption{Left panel: Correlation between the quadrupole moment and the squared charge radius (dashed line). The red solid line indicates our results in case of $r_1$ within the estimated interval of $\left[-150,-50\right]$ MeV. Right panel: Correlation between the squared quadrupole radius and the neutron separation energy (dashed line). The red cross indicates our result for $S_n=0.15$ MeV. The blue shaded bands give our estimate of the EFT uncertainty.}
    \label{fig:Correlations}
  \end{figure}
%%%%%%%%%%%%%%%%%%%%%%%%%%%%%%%%%%%%%%%%%%%%%%%%%%%%%%%%%%%%%%%%%%%%%%%%
  The electromagnetic form factors from the previous section can be expanded for low three-momentum transfer $q^{2}$ in order to extract the
  corresponding radii:
	\begin{align}
			G_{\text{(E/M)}L}(q)&= 1-\frac{1}{6}\langle r_{\text{(E/M)}L}^{2} \rangle q^{2}+...,
	\end{align}
where $\langle r_{\text{(E/M)}L}^{2}\rangle$ denotes the expectation value of the electric/magnetic radius squared with multipolarity $L$, respectively.
Below, we give general expressions for a halo nucleus with a $J^P=0^+$ core
and a halo neutron in a $J^P=3/2^-$ $P$-wave state as well as explicit
numbers for $^{31}$Ne based on the assumptions discussed in
Sec.~\ref{sec:HaloEFT}.

\subsection{Results in the Electric Sector}
The electric monopole and quadrupole radii at LO are
	\begin{align}
		\label{eq:MonopoleRadius}
		\langle r_{\text{E}0}^{2} \rangle &=-\frac{5y^{2}}{2\gamma_1 r_1},\\
		\label{eq:QuadrupoleRadius}
		\langle r_{\text{E}2}^{2} \rangle &=\frac{3y^{2}}{5\gamma_1^{2}}.
	\end{align}
        Given these expressions, we can establish universal correlations with other observables.
        Considering the result for the quadrupole moment in \cref{eq:QuadrupoleMoment}, we find the  correlation
	\begin{align}
		\mu_Q=\frac{2}{15}q_c \langle r_{\text{E}0}^{2}\rangle,
	\end{align}
        which is depicted in the left panel of \cref{fig:Correlations}. The red solid line indicates our result for $^{31}$Ne
        for $r_1$ within the estimated interval of $\left[-150,-50\right]$ MeV. We obtain
	\begin{align}
		\sqrt{\langle r_{\text{E}0}^{2}\rangle}&\in \left[0.20(08),\,0.35(14)\right]\text{ fm}\,,\\
		\mu_Q&\in \left[0.06(02),\,0.17(07)\right]\text{ fm}^{2}\,,
	\end{align}
where the numbers in parentheses give the EFT uncertainties of $40$\%.

Furthermore, we find a correlation between the quadrupole radius squared $\langle r_{\text{E}2}^{2}\rangle$ and
the neutron separation energy $S_n$ given by
	\begin{align}
		\label{eq:CorrelationrSn}
		\langle r_{\text{E}2}^{2}\rangle=\frac{3y^{2}}{10m_R}\frac{1}{S_n}.
	\end{align}
It is depicted in the right panel of \cref{fig:Correlations}, where the red cross indicates our $^{31}$Ne result for $S_n=0.15$ MeV given by 
	\begin{align}
		\sqrt{\langle r_{\text{E}2}^{2}\rangle}=0.30(12)\text{ fm}.
	\end{align}

\subsection{Results in the Magnetic Sector}
The LO result for the magnetic octupole radius squared reads
	\begin{align}
		\langle r_{\text{M}3}^{2}\rangle&=\frac{3(1-y)^{2}}{5\gamma_1^{2}}=\frac{3(1-y)^{2}}{10\mu}\frac{1}{S_n}.
	\end{align}
This correlation between the squared octupole radius and the neutron separation energy is similar to the correlation in \cref{eq:CorrelationrSn}. The corresponding
value for the octupole radius of $^{31}$Ne is
	\begin{align}
		\sqrt{\langle r_{\text{M}3}^{2}\rangle}=9.0(3.6)\text{ fm}\,.
	\end{align}
Given the octupole moment in \cref{eq:OctupoleMoment} and estimating $r_1$ as before yields the following result for the octupole moment of $^{31}$Ne
	\begin{align}
		\mu_O\in\left[-14(6),-5(2)\right] \text{ $\mu_N$fm}^{2}.
	\end{align}
        Since the magnetic dipole moment contains the counterterm proportional to $L_{\text{M}}(\mu)$, it is not possible to predict its value in Halo EFT.
        However, as discussed above, the full $q^2$-dependence of
        $\mu_D G_{\text{M}1}(q)$ is predicted. In particular,
        the ``renormalized magnetic radius'' defined as
        $\mu_D \langle r_{\text{M}1}^{2}\rangle$ is independent of $L_{\text{M}}(\mu)$. For a \mbox{$J=3/2$} state, we have
\begin{align}
  \mu_D \langle r_{\text{M}1}^{2}\rangle &= -\frac{\mu_N}{\gamma_1 r_1} \left( \frac{3 y^3 q_c}{2A_c} + \frac{27}{10}(1-y)^2 \kappa_n \right)\,.
\end{align}
Using this relation, it is either possible to predict $\langle r_{\text{M}1}^{2}\rangle$ once the magnetic dipole moment is determined experimentally or vice versa.
The corresponding result for \mbox{$J=1/2$} can again be found in \mbox{Appendix \ref{sec:AppJ12}}. 
\subsection{Nuclear Deformation\label{sec:Deformation}}
The appearance of higher multipole moments such as the electric quadrupole as well as the magnetic octupole moment indicates that $^{31}$Ne is not a spherically symmetric
nucleus. Following Ref.~\cite{Zelevinskych12},
we assume a quadrupolar deformed shape with a sharp edge at radius
	\begin{align}
		R_{\text{def}}=R_0\left( 1+\beta_2 Y_{20}(\theta,\phi) \right)/N,
	\end{align}	
where $R_0$ is the equilibrium radius, meaning the radius if the nucleus would be spherically symmetric. The additional term $\beta_2 Y_{20}(\theta,\phi)$ accounts for the quadrupolar 
deformation where $\beta_2$ is called the deformation parameter. Having defined this surface radius and using $\beta_2\ll1$, we can relate it to the spectroscopic quadrupole moment via~\cite{Zelevinskych12,greiner1996nuclear}
	\begin{equation}
		\begin{aligned}
			\label{eq:deformation}
			\mu_Q\left(3/2\right)&=\frac{1}{5}\sqrt{\frac{16\pi}{5}}\frac{3}{4\pi} Z e R_0^{2} \beta_2\\
			&=\sqrt{\frac{1}{5\pi}} Z e \beta_2 \langle r_{\text{E}0}^{2}\rangle.
		\end{aligned}
	\end{equation}
        In the second line of \cref{eq:deformation} we used $\langle r_{\text{E}0}^{2}\rangle=(3/5)R_0^{2}$. As a result, we find a linear correlation between the quadrupole moment and the mean squared electric monopole radius. This is exactly the same correlation we found in our Halo EFT calculation and hence equating the proportionality factors allows us to determine the deformation parameter of $^{31}$Ne to be $\beta_2=0.53$. This value is similar to $\beta_2=0.41$ found
        in an antisymmetrized molecular dynamics calculation with the Gogny D1S
        interaction~\cite{PhysRevLett.108.052503}.   
        A deformation parameter of $\beta_2\approx0.4$ was also obtained in
Ref.~\cite{SHUBHCHINTAK201499} from the analysis of
parallel momentum distribution of the charged fragment in the breakup
of $^{31}$Ne. 
These values indicate a significant deformation due to the non-vanishing quadrupole moment. However, we note that our prediction is solely determined by the dynamics of the electrically charged core. The deformation of the core itself is not included here. In this sense, our predictions are relative to the core. Once the intrinsic properties of the core are experimentally determined, they can be included in our theory. In particular, intrinsic deformation properties of the core such as its quadrupole moment due to the $J^P=2^+$ excited state can be described explicitly in Halo EFT by including a corresponding field in the effective Lagrangian. This would allow us to predict the deformation properties due to both the intrinsic core properties and the dynamics of the halo nucleus. Indeed, it is expected to find a quadrupolar deformation of the $^{30}$Ne core. Urata et al.~\cite{PhysRevC.83.041303} showed that the deformation parameter of the $^{30}$Ne core is around $\beta_2=0.2...0.3$,
        while Minomo et al.~\cite{PhysRevLett.108.052503} found
        $\beta_2=0.39$.      
Therefore, the total quadrupolar deformation of $^{31}$Ne is ultimately composed of both deformation effects.

\section{E1 Breakup: \texorpdfstring{$^{\mathbf{31}}$Ne}{31Ne} into \texorpdfstring{$^{\mathbf{30}}$Ne}{30Ne} and a Neutron\label{sec:E1}}
\begin{figure}[t]
  \centering
  \includegraphics[width = 0.25\textwidth]{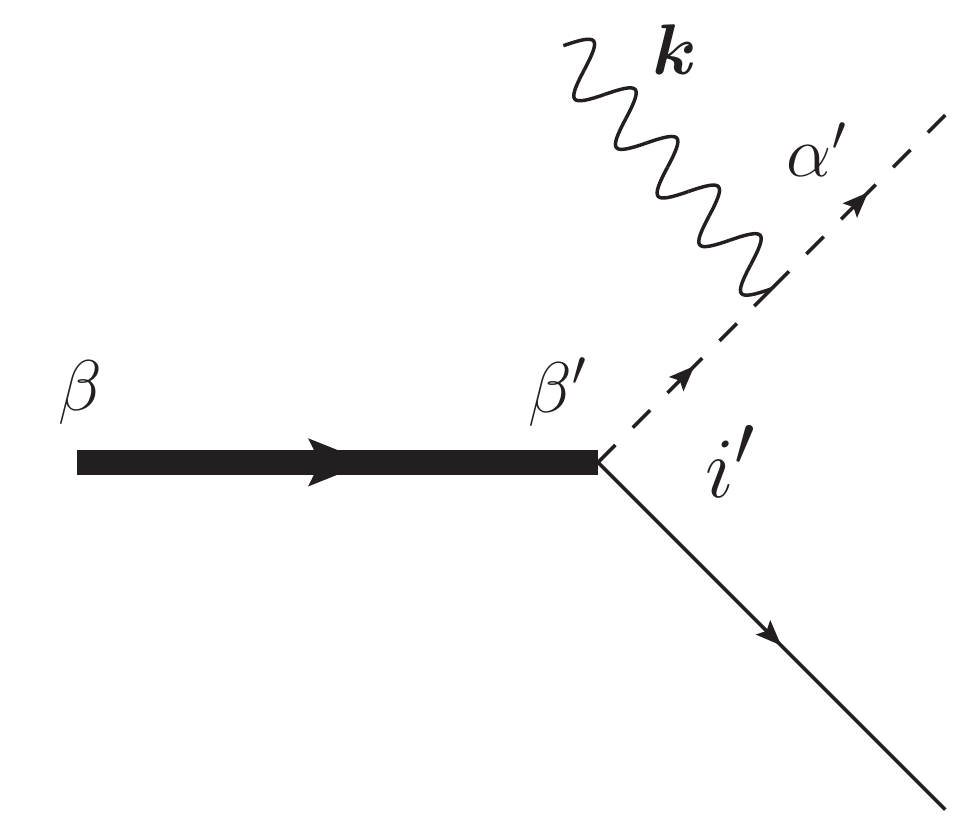}
  \caption{\label{fig:E1}E1 Breakup of $^{31}$Ne into the continuum consisting of the core and neutron. We use the same notation as in \cref{fig:SA31Ne}.}
\end{figure}

In \cref{fig:E1}, we show the LO diagram contributing to the E1 breakup of $^{31}$Ne. The photon transfers an angular momentum of $1$ onto the two-body system consisting of the core and neutron. Since this two-body system is bound in a $P$-wave, the possible final angular momenta in the continuum are $0$ and $2$, corresponding to an $S$- and a $D$-wave, respectively.
%%%%%%%%%%%%%%%%%%%%%%%%%%%%%%%%%%%%%%%%%%%%%%%%%%%%%%%%%%%%%%%%%%
%%%%%%%%%%%%%%%%%%%%%%%%%%%%%%%%%%%%%%%%%%%%%%%%%%%%%%%%%%%%%%%%%

The scalar transition amplitude in momentum space is given by
\begin{align}
  \Gamma_0^{\alpha' \beta}&=\frac{i2m_R g_1 q_c e \sqrt{Z_\pi} C_{\scriptscriptstyle{(1i')(\frac{1}{2}\alpha')}}
										^{\scriptscriptstyle{\frac{3}{2}\beta}} (\bs{p}-y\bs{k})_{i'}}{\gamma_1^2+(\bs{p}-y\bs{k})^2}\,,\\
										&=iq_c e \braket{\bs{p}-y{\bs{k}}|\psi^{\alpha'\beta}}\label{eq:WFOverlap}\,,
\end{align}
where $\ket{\psi^{\alpha'\beta}}$ represents the bound state of $^{31}$Ne (see Appendix \ref{sec:AppPWaveWF} for explicit expressions),
the mass ratio $y$ is defined in \cref{eq:defy},
$\bs{p}$ is the relative momentum between the core and the neutron, while $\bs{k}$ is the photon momentum.

Without loss of generality, we choose the photon to be traveling in the $\hat{z}$-direction.
We insert in \cref{eq:WFOverlap} an identity operator in configuration space and express $e^{iykz}$ by its plane wave expansion
\begin{align}
  e^{iykz}=e^{iykr\cos \theta}=\sqrt{4\pi} \sum_L \sqrt{2L+1} i^L j_L(ykr) Y_{L0}(\mb{e}_{\mb{r}})\,,
\end{align}
where $j_L(x)$ is a spherical Bessel function. In the low-energy limit, we use $j_L(ykr)\approx(ykr)^L/(2L+1)!!$.

The scalar transition amplitude then reads
\begin{align}
  \Gamma_0^{\alpha' \beta}=i \sum_L \int \text{d}^3r \bra{\bs{p}} \hat{\tilde{\rho}}_L(\bs{r})\ket{\psi^{\alpha'\beta}} \sqrt{4\pi} \sqrt{2L+1} i^L \frac{(kr)^L}{(2L+1)!!} Y_{L0}(\mb{e}_{\mb{r}})\,,
\end{align}
where 
\begin{align}
  \hat{\tilde{\rho}}_L(\bs{r}) = e Z_{\text{eff}}^{(L)} \ket{\bs{r}}\bra{\bs{r}}
\end{align}
with $Z_{\text{eff}}^{(L)}=q_c y^L$.

This means, that the photon in \cref{fig:E1} transfers all possible angular momenta.
For a specific angular momentum transfer, the amplitude reads
\begin{align}
  \Gamma_0^{\alpha' \beta}(\text{EL}; M=0)=i \int \text{d}^3r \bra{\bs{p}} \hat{\tilde{\rho}}_L(\bs{r})\ket{\psi^{\alpha'\beta}} \sqrt{4\pi} \sqrt{2L+1} i^L \frac{(kr)^L}{(2L+1)!!} Y_{L0}(\mb{e}_{\mb{r}})\,.
\end{align}
The matrix element relevant for the calculation of the EL breakup is given by~\cite{GreinerMaruhn} 
\begin{align}
  {\mathcal{M}}(\text{EL};0)&=\int \text{d}^3r \bra{\bs{p}}\hat{\tilde{\rho}}_L(\bs{r})\ket{\psi^{\alpha'\beta}} r^L Y_{L0}(\mb{e}_{\mb{r}})\,,\\
  &= -i \frac{(2L+1)!!}{k^L} (i)^{-L} \sqrt{\frac{1}{4\pi (2L+1)}} \Gamma_0^{\alpha' \beta}(\text{EL}; 0)\label{eq:MEL}\,.
\end{align}

Since we are interested in the E1 breakup, we set $L=1$ in \cref{eq:MEL}. Moreover, we plug in the bound state wave function in configuration space given by
\begin{equation}
\begin{aligned}
  \braket{\bs{r}|\psi^{\alpha'\beta}}&=i \sqrt{\frac{-2\gamma_1^2}{r_1}} \frac{e^{-\gamma_1r}}{r}\left( 1+\frac{1}{\gamma_1r} \right) Y_{1i'}(\mb{e}_{\mb{r}}) C_{\scriptscriptstyle{(1i')(\frac{1}{2}\alpha')}}
										^{\scriptscriptstyle{\frac{3}{2}\beta}}\\
										&=i\,C_{\pi} \frac{u(r)}{r} Y_{1i'}(\mb{e}_{\mb{r}})C_{\scriptscriptstyle{(1i')(\frac{1}{2}\alpha')}}^{\scriptscriptstyle{\frac{3}{2}\beta}}\,,
\end{aligned}
\end{equation}
with $C_{\pi}$ denoting the asymptotic normalization constant (ANC) and $u(r)$ the radial wave function. They read
\begin{align}
  C_{\pi}&=\sqrt{\frac{-2\gamma_1^2}{r_1}}\,,\\
  u(r)&= e^{-\gamma_1r}\left( 1+\frac{1}{\gamma_1r} \right)\,.
\end{align}
This yields
\begin{align}
  {\mathcal{M}}(\text{E1};0)&= ieZ_{\text{eff}}^{(1)} C_{\pi} \int \text{d}^3r e^{-i\bs{p}\bs{r}} u(r) Y_{1i'}(\mb{e}_{\mb{r}}) Y_{10}(\mb{e}_{\mb{r}}) C_{\scriptscriptstyle{(1i')(\frac{1}{2}\alpha')}}^{\scriptscriptstyle{\frac{3}{2}\beta}}\label{eq:ME1}\,.
\end{align}
The product of the spherical harmonics in \cref{eq:ME1} can be expressed as an irreducible sum of spherical harmonics with $L=0$ and $L=2$. This allows us to extract the two relevant matrix elements for a transition into either an $S$-wave or a $D$-wave.

The $P\rightarrow S$ transition amplitude reads
\begin{align}
  {\mathcal{M}}(\text{E1};0;P\rightarrow S)=\frac{ieZ_{\text{eff}}^{(1)} C_{\pi}}{\sqrt{4\pi}} \int \text{d}^3r e^{-i\bs{p}\bs{r}} u(r) Y_{00}(\mb{e}_{\mb{r}}) C_{\scriptscriptstyle{(10)(\frac{1}{2}\alpha')}}^{\scriptscriptstyle{\frac{3}{2}\beta}}
\end{align}
We use the plane wave expansion of $e^{-i\bs{p}\bs{r}}$, integrate over $d\Omega$ and couple the angular momentum of $L=0$ and the spin of the neutron to a two-particle continuum with total spin quantum numbers $(J'\beta')$ to find
\begin{align}
  {\mathcal{M}}^{J'\beta'\beta}(\text{E1};0;P\rightarrow S)=ieZ_{\text{eff}}^{(1)} C_{\pi} \sqrt{4\pi} Y_{00}({\bs{e}_{\bs{p}}}) \int \text{d}r j_0(pr) u(r) r^2 C_{\scriptscriptstyle{(10)(\frac{1}{2}\alpha')}}^{\scriptscriptstyle{\frac{3}{2}\beta}} C_{\scriptscriptstyle{(00)(\frac{1}{2}\alpha')}}^{\scriptscriptstyle{J'\beta'}}
\end{align}
We proceed similarly for the $P \rightarrow D$ transition amplitude and find
\begin{equation}
\begin{aligned}
  {\mathcal{M}}^{J'\beta'\beta}(\text{E1};0;P\rightarrow D)=&-ieZ_{\text{eff}}^{(1)} C_{\pi} \sqrt{\frac{24\pi}{5}} \sum_m Y_{2m}(\bs{e}_{\bs{p}}) \\
 &\times \int \text{d}r j_2(pr) u(r) r^2  C_{\scriptscriptstyle{(1i')(10)}}^{\scriptscriptstyle{2m}} C_{\scriptscriptstyle{(1i')(\frac{1}{2}\alpha')}}^{\scriptscriptstyle{\frac{3}{2}\beta}} C_{\scriptscriptstyle{(2m)(\frac{1}{2}\alpha')}}^{\scriptscriptstyle{J'\beta'}}\,.
\end{aligned}
\end{equation}
Performing the radial integral and sum (average) over final (initial) spins and multiply our results with a factor of $3$ to make up for the fact that we chose the photon to propagate in the $\hat{z}$-direction, we finally get
\begin{align}
  \overline{\left| {\mathcal{M}}(\text{E1}; P \rightarrow S) \right|}^2 &= e^2 {Z_{\text{eff}}^{(1)}}^2 \left| C_{\pi} \right|^2 4\pi \left| Y_{00}(\bs{e}_{\bs{p}}) \right|^2 \frac{(3\gamma_1^2+p^2)^2}{\gamma_1^2(\gamma_1^2+p^2)^4}\,,\\
  \overline{\left| {\mathcal{M}}(\text{E1}; P \rightarrow D) \right|}^2 &= e^2 {Z_{\text{eff}}^{(1)}}^2 \left| C_{\pi} \right|^2 \frac{24\pi}{5} \sum_m \left| Y_{2m}(\bs{e}_{\bs{p}}) \right|^2 \frac{4p^4}{\gamma_1^2(\gamma_1^2+p^2)^4} \sum_{i'}\left(C_{\scriptscriptstyle{(1i')(10)}}^{\scriptscriptstyle{2m}}\right)^2 \,.
\end{align}
%%%%%%%%%%%%%%%%%%%%%%%%%%%%%%%%%%%%%%%%%%%%%%%%%%%%%%%%%%%%%%%%%%%%%%%
\begin{figure}[t]
  \centering
  \includegraphics[width = 0.49\textwidth]{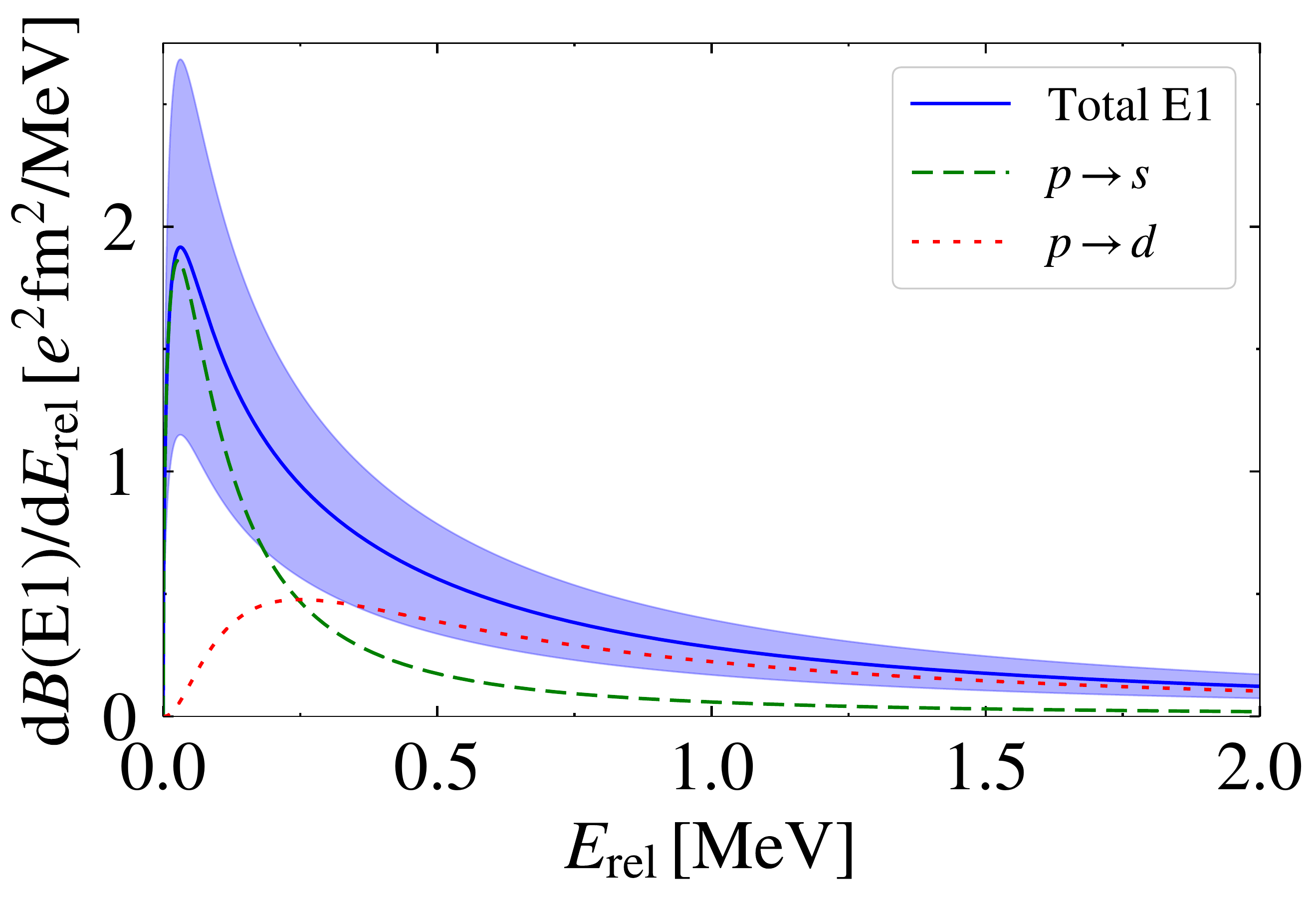}
  \includegraphics[width = 0.49\textwidth]{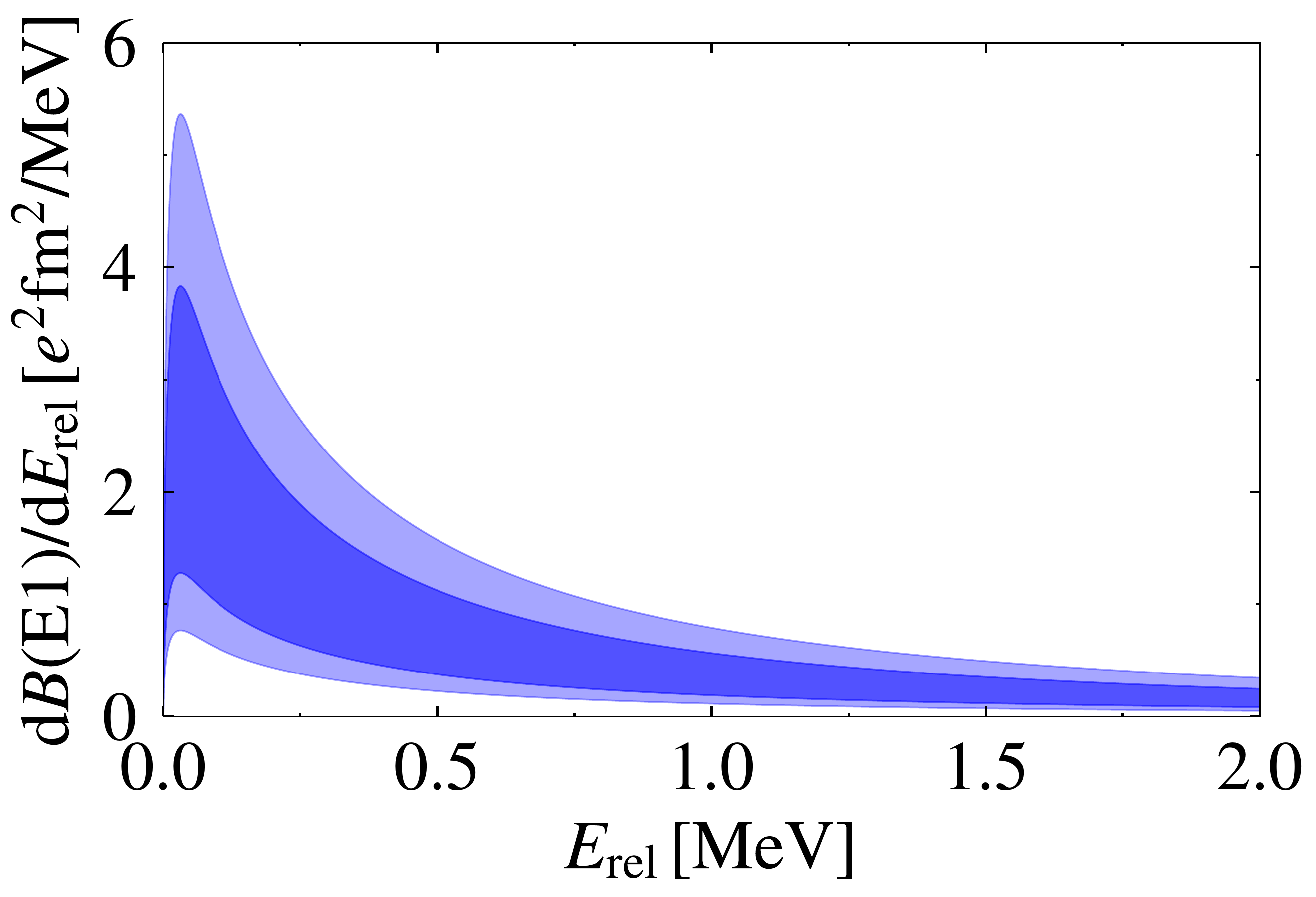}
  \caption{\label{fig:diffBE1}Differential E1 transition strength of $^{31}$Ne as a function of the relative energy $E_{\text{rel}}$ of the $^{30}$Ne core and the neutron. Left panel: Separate contributions shown for $r_1=-100$ MeV. Explanation of curves is given in inset. Right panel: Dark shaded band gives the uncertainty from $r_1$ estimated
    as $r_1 \in [-150, -50]$~MeV, while the light shaded bands also
    include the general EFT uncertainty at LO of 40\%}
\end{figure}
%%%%%%%%%%%%%%%%%%%%%%%%%%%%%%%%%%%%%%%%%%%%%%%%%%%%%%%%%%%%%%%%%%%%%%
The differential E1 transition strength is given by~\cite{Typel:2004us}
\begin{align}
  \text{dB}(\text{E1})= \left| {\mathcal{M}}(\text{E1}) \right|^2 \frac{\text{d}^3p}{(2\pi)^3}\,.
\end{align}
Using $E_{\text{rel}}=p^2/(2m_R)$, the differential E1 transition strength as a function of the relative energy $E_{\text{rel}}$ between the core and neutron reads
\begin{align}
  \frac{\text{dB}(\text{E1})}{\text{d}E_{\text{rel}}}=\frac{1}{(2\pi)^3}\int \text{d}\Omega_{\bs{p}} m_R \, p \left| {\mathcal{M}}(\text{E1}) \right|^2\,.
\end{align}
The final results for both transitions read
\begin{align}
  \frac{\text{dB}(\text{E1};P \rightarrow S)}{\text{d}E_{\text{rel}}}&= e^2 {Z_{\text{eff}}^{(1)}}^2 \frac{m_R}{2\pi^2} \left| C_{\pi} \right|^2 \frac{p}{\gamma_1^2} \frac{(3\gamma_1^2+p^2)^2}{(\gamma_1^2+p^2)^4}\,,\\
  \frac{\text{dB}(\text{E1};P \rightarrow D)}{\text{d}E_{\text{rel}}}&= e^2 {Z_{\text{eff}}^{(1)}}^2 \frac{m_R}{\pi^2} \left| C_{\pi} \right|^2 \frac{p}{\gamma_1^2} \frac{4p^4}{(\gamma_1^2+p^2)^4}\,.
\end{align}
We show the corresponding curves in \cref{fig:diffBE1} for a $P$-wave effective momentum \mbox{$r_1=-100~\text{MeV}$}. The total differential B(E1) transition strength is given in blue while the blue shaded band represents an 40\% estimate
of the EFT uncertainty at LO from our power counting. As expected,
the $S$-wave contribution in the continuum dominates at low energies while
the $D$-wave contribution takes over around $E_{\text{rel}}=0.25$~MeV.

\begin{figure}[t]
  \centering
  \includegraphics[width = 0.7\textwidth]{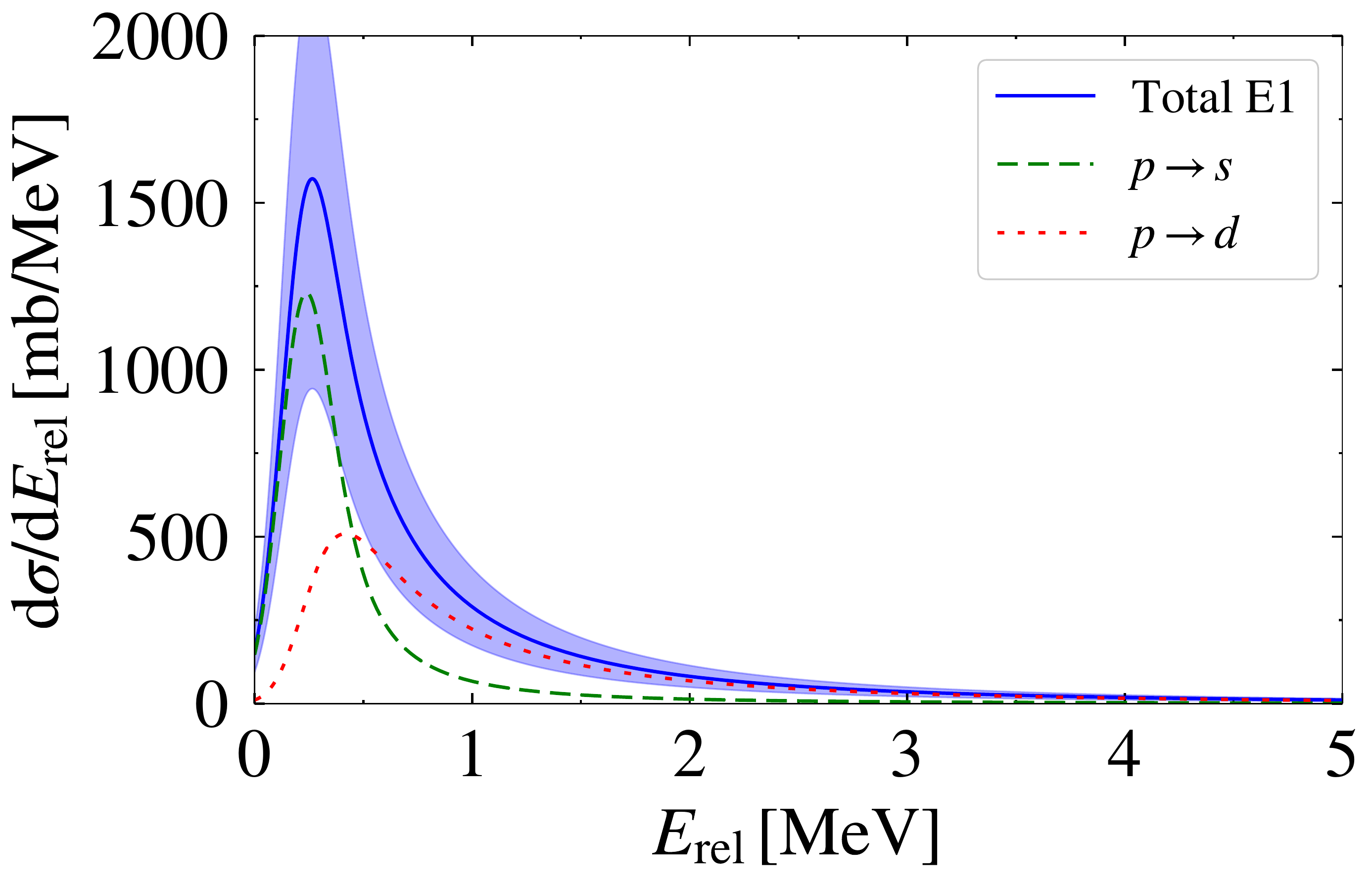}
  \caption{\label{fig:diffcross}Differential cross section for E1 breakup of $^{31}$Ne folded with a Gaussian energy resolution of width $0.1~$MeV
    as a function of the relative energy $E_{\text{rel}}$ of the $^{30}$Ne core and the neutron for $r_1=-100~$MeV. The light shaded band gives the EFT uncertainty at LO of $40$\%. Explanation of curves is given in inset.} 
\end{figure}

\begin{figure}[t]
  \centering
  \includegraphics[width = 0.7\textwidth]{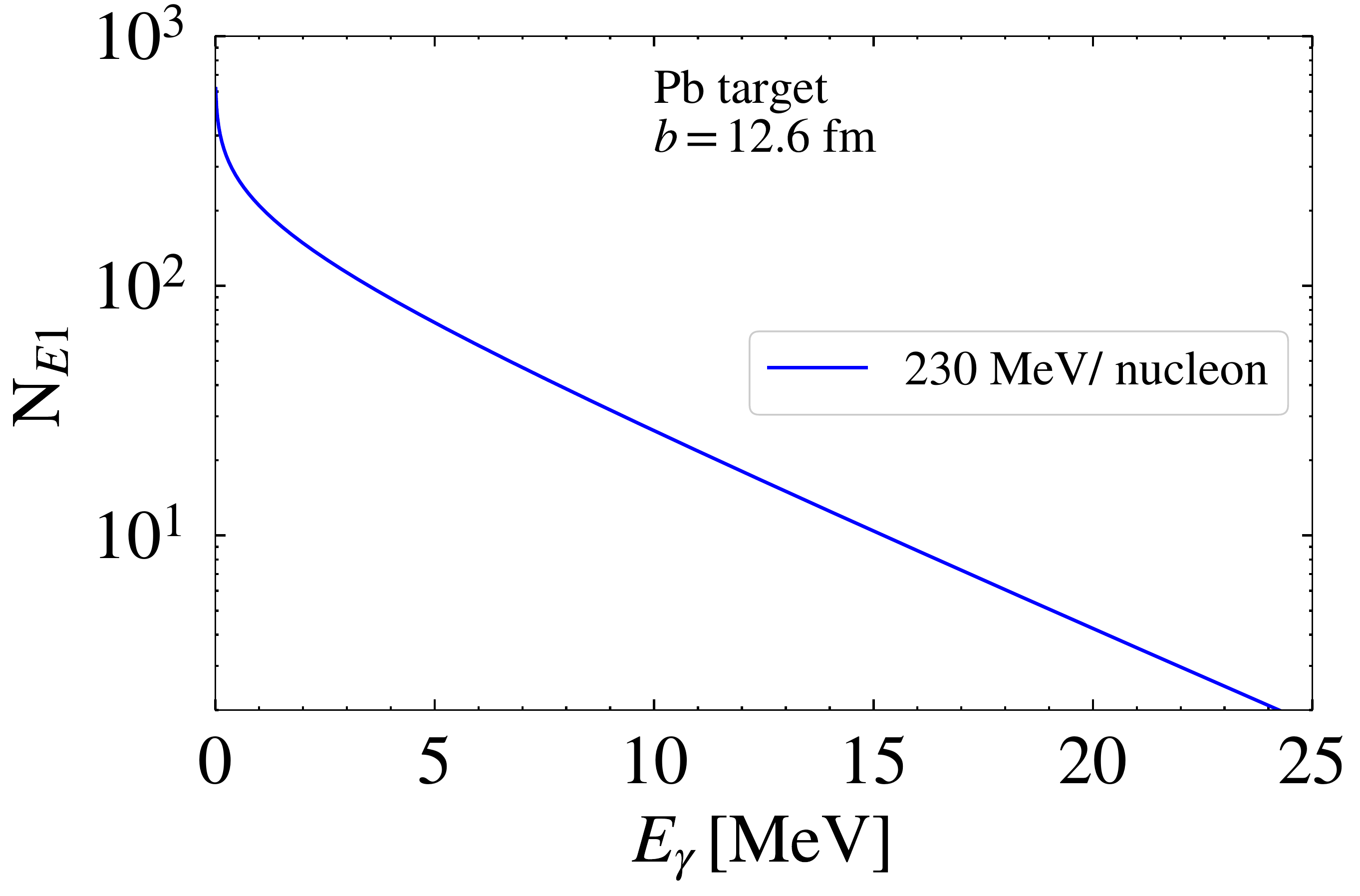}
  \caption{\label{fig:virtualphoton}Virtual photon number as a function of the photon energy $E_{\gamma}$ for a lead target with an impact parameter of $b=12.6$~fm and a kinetic energy of $230~$MeV per nucleon~\cite{PhysRevLett.103.262501} calculated using the expression given in
   Ref.~\cite{Bertulani:2009zk}.} 
\end{figure}

In Fig.~\ref{fig:diffcross}, we show the differential cross section
for the E1 breakup of $^{31}$Ne.
%%%%%%%%%%%%%%%%%%%%%%%%%%%%%%%%%%%%%%%%%%%%%%%%%%%%%%%%%%%%%%%%%%%%%%%
%%%%%%%%%%%%%%%%%%%%%%%%%%%%%%%%%%%%%%%%%%%%%%%%%%%%%%%%%%%%%%%%%%%%%%
The bare cross section corresponding to an infinite energy resolution of the
detector is given by
\begin{equation}
  \frac{d\sigma}{dE_{\text{rel}}}(E_{\text{rel}}) =
\frac{16\pi^3}{9}N_\text{E1}(E_\gamma)\frac{\text{dB}(\text{E1})}{\text{d}E_{\text{rel}}},
\end{equation}
where $E_\text{rel}=E_\gamma -S_n + k^2/(2M_{nc})$ with $S_n$ the neutron separation
energy of $^{31}$Ne and
the expression for the virtual photon flux $N_\text{E1}$ is taken from
Ref.~\cite{Bertulani:2009zk}. The corresponding plot can be found in Fig.~\ref{fig:virtualphoton}.
%%%%%%%%%%%%%%%%%%%%%%%%%%%%%%%%%%%%%%%%%%%%%%%%%%%%%%%%%%%%%%%%%%%%%%%
%%%%%%%%%%%%%%%%%%%%%%%%%%%%%%%%%%%%%%%%%%%%%%%%%%%%%%%%%%%%%%%%%%%%%%
Since no experimental results
have been published to date, we have folded the predicted bare cross section
with a hypothetical Gaussian energy resolution,
\begin{equation}
  g(E)=\frac{1}{\sqrt{2\pi}\,\sigma} e^{-\frac{1}{2}(E/\sigma)^2},
\end{equation}
with a constant width $\sigma=0.1$~MeV to allow for a more realistic
comparison to future experimental data.
The resolution averaged cross section is obtained from
\begin{equation}
  \left\langle\frac{d\sigma}{dE_{\text{rel}}}(E_{\text{rel}})\right\rangle =
  \int dE' \,\frac{d\sigma}{dE_{\text{rel}}}(E')\, g(E_{\text{rel}}-E'),
\end{equation}
and shown in Fig.~\ref{fig:diffcross}.
The dark shaded band gives the uncertainty from $r_1$ estimated
as $r_1 \in [-150, -50]$~MeV, while the light shaded bands also
includes the general EFT uncertainty at LO of 40\%.
The absolute height of the peak has a large EFT uncertainty
while the shape of the curve is rather robust. However, it
is strongly influencend
by the assumed detector resolution parameter $\sigma$ and larger values
of $\sigma$ will further smear the peak.

\section{Conclusion\label{sec:Conclusion}}
We have investigated the electromagnetic properties of $^{31}$Ne using Halo EFT. Instead of using standard Cartesian coordinates, we have introduced a spherical basis
that is ideally suited for the description of halo nuclei beyond the $S$-wave. It uses the 
correct number of degrees of freedom by construction and therefore leads to more compact and simplified
expressions. We expect it to be useful in future studies of halo nuclei
with higher angular momentum.

In our study of the electric properties, we found that our numerical predictions are fairly small. We calculated
the charge radius as $\langle r_{\text{E}0}^{2}\rangle^{1/2} \in \left[0.20(08),0.35(14)\right]~\text{fm}$ and
the quadrupole moment as $\mu_Q \in \left[0.06(02),0.17(07)\right]~\text{fm}^{2}$, where the numbers in square brackets give the results for
$r_1 =-150$~MeV and $-50$~MeV, respectively, while the number in
parentheses indicates the error from higher orders in the EFT expansion.
The corresponding quadrupole radius is independent of $r_1$ and
predicted to be $\langle r_{\text{E}2}^{2}\rangle^{1/2}=0.30(12)~\text{fm}$.

These values are rather small because at leading order they are solely
determined by the motion of the electrically charged core around the center
of mass.  Since the $^{30}$Ne core is almost
as heavy as the total system, $^{31}$Ne, this yields small predictions.
Therefore, we expect internal electric properties of the core to be important,
at least in the electric sector. Such effects can be included at NLO by
treating excited states of the core as explicit fields or via counterterms.
See Ref.~\cite{Ryberg:2019cvj} for a discussion of this issue in the case
of the charge radius.

Once more experimental data is available, the treatment of
the first excited state of the $^{30}$Ne core as an explicit degree of freedom within Halo EFT becomes feasible. This would lead to more precise predictions in the electric sector and potentially to a smaller expansion parameter.
Alternatively, one could include the excited states of the core by
describing the halo nucleus as a neutron coupled to a rotor similar
to the work of Refs.~\cite{Papenbrock:2020zhh,Alnamlah:2020cko}.

Nevertheless, in the magnetic sector the main contribution
to observables arises from the motion of the valence neutron
around the center of mass. This means that corrections due to the internal core properties at NLO should be negligible. As a matter of fact, in the magnetic case our numerical predictions are much larger,
$\langle r_{\text{M}3}^{2}\rangle^{1/2}=9.0(3.6)~\text{fm}$
and $\mu_{O}\in\left[-14(6),-5(2)\right] \text{ $\mu_N$fm}^{2}$. The large octupole radius reveals the size of the halo system. 
Unfortunately, the magnetic dipole moment cannot be predicted since it depends on the counterterm $L_{\text{M}}$ already at LO.

In general, the non-vanishing higher multipole moments with multipolarity $L>1$ indicate that $^{31}$Ne is not a spherically symmetric nucleus. We extracted the $\beta_2$-deformation parameter from the linear correlation between the quadrupole moment and the charge radius and found $\beta_2=0.53$. This value indicates a significant deformation due to the quadrupole moment. However, we note that this prediction is solely determined by the dynamics of the electrically charged core whereas the deformation of the core itself is not included here. In this sense, our prediction is relative to the core. Intrinsic deformation properties of the core can be described explicitly in Halo EFT by introducing corresponding fields in the effective Lagrangian. Once more experimental data of the intrinsic deformation properties of the core such as its quadrupole moment are available, they can be included in Halo EFT to calculate the total deformation properties due to both the intrinsic core properties and the dynamics of the halo nucleus.

Moreover, we have derived the differential B(E1) transition strength as a function of the relative energy $E_{\text{rel}}$ between the $^{30}$Ne core and the neutron. This transition strength together with the virtual photon number allowed us to calculate the differential cross section for E1 breakup of $^{31}$Ne. In order to take into account a realistic limited energy resolution in experiment, we have folded our results with a Gaussian energy resolution of width $0.1$ MeV. Comparing these results to future data will help us to further determine unknown parameters which in turn enables us to improve our Halo EFT for $^{31}$Ne.  Finally, an application of our formalism to $^{37}$Mg~\cite{Kobayashi:2014owa},
which is also a candidate for a deformed $P$-wave halo nucleus appears promising.

\section*{Acknowledgements}
This work was supported by the Deutsche Forschungsgemeinschaft (DFG, German
Research Foundation) -- Projektnummer 279384907 -- CRC 1245
and by the German Federal Ministry of Education and Research (BMBF) (Grant
no. 05P21RDFNB).

\newpage
\appendix
\section{Spherical Coordinates}
\label{sec:AppSpherical}
Throughout this work, we use spherical coordinates. The components of a three-dimensional vector are written as
	\begin{align}
		r_i&=\sqrt{\frac{4\pi}{3}} r Y_{1i}(\mb{e}_{\mb{r}})\,,\\
		r^i&=\sqrt{\frac{4\pi}{3}} r Y_{1i}^*(\mb{e}_{\mb{r}})\,,
	\end{align}
with the connection to Cartesian coordinates given by
	\begin{align}
		r_1&=\left(r^{1}\right)^*=\sqrt{\frac{4\pi}{3}}r Y_{11}=\sqrt{\frac{4\pi}{3}}r \left(-\sqrt{\frac{3}{8\pi}}\right)\sin(\theta)e^{i\phi}=-\frac{r_x+ir_y}{\sqrt{2}},\\
		r_0&=\left(r^{0}\right)^*=\sqrt{\frac{4\pi}{3}}r Y_{10}=\sqrt{\frac{4\pi}{3}}r \sqrt{\frac{3}{4\pi}}\cos(\theta)=r_z,\\
		r_{-1}&=\left(r^{-1}\right)^*=\sqrt{\frac{4\pi}{3}}r Y_{1-1}=\sqrt{\frac{4\pi}{3}}r \sqrt{\frac{3}{8\pi}}\sin(\theta)e^{-i\phi}=\frac{r_x-ir_y}{\sqrt{2}}.
	\end{align}

The components of the nabla operator in spherical coordinates expressed in terms of the Cartesian components are given by
  \begin{align}
    \nabla_1&=-\frac{1}{\sqrt{2}}\left(\frac{\partial}{\partial x}+i\frac{\partial}{\partial y}\right)\,,\\
    \nabla_0&=\frac{\partial}{\partial z}\,,\\
    \nabla_{-1}&=\frac{1}{\sqrt{2}}\left(\frac{\partial}{\partial x}-i\frac{\partial}{\partial y}\right)\,.
  \end{align}

\section{\textit{J}=1/2 Case}
\label{sec:AppJ12}
The on-shell neutron-core scattering amplitude for a $P$-wave with $J=1/2$ in the center-of-mass frame with $E=p^2/(2m_R)=p'^2/(2m_R)$ and $p=|\mb{p}|=|\mb{p}'|$ reads
\begin{align}
		T_{\alpha^{\prime}\alpha}(\mb{p}',\mb{p})	=\frac{6\pi}{m_R}~\frac{\frac{1}{3}\,\mb{p}'\cdot\mb{p}\,\delta_{\alpha'\alpha}+\frac{i}{3}
		\Big(\mb{\sigma} \cdot \left(\mb{p}'\times\mb{p}\right)\Big)_{\hspace{-0.1cm}\alpha'\alpha}}
		{\left(\frac{6\pi\Delta_1}{m_R g_1^2}+\frac{3\pi\eta_1}{m_R^2 g_1^2}p^2+\frac{3}{2}\mu p^2+ip^3\right)}\,,
	\end{align}
where $\mb{\sigma}$ is the three-dimensional vector with the Pauli matrices as its components.

The corresponding LO result for the magnetic dipole form factor for a $J=1/2$ dimer is
	\begin{align}
		G_{\text{M}1}(q)&=\frac{1}{r_1}\left[ \frac{a L_{\text{M}}^{1/2}(\mu) + b \kappa_n + c \frac{yq_c}{A_c} }{\mu_D} \right]\mu_N,\\
	\end{align}
with $A_c$ denoting the mass number of the core, while $a,b$ and $c$ are functions given by
	\begin{align}
		a&=\left[r_1+3\left(\mu-\gamma_1\right)\right],\\
		b&=\mu-\gamma_1-\frac{(1-y)q}{2}\arctan\left(\frac{(1-y)q}{2\gamma_1}\right),\\
		c&=\frac{1}{2q}\left[2\gamma_1 q+\frac{\left(4\gamma_1^{2}+y^{2}q^{2}\right)}{y}\arctan\left(\frac{yq}{2\gamma_1}\right)\right]-2\mu\,,
	\end{align}
where the definition of the mass ratio $y$ is given in \cref{eq:defy}.        
The dipole moment of a $J=1/2$ dimer also depends on an unknown
counterterm $L_{\text{M}}^{1/2}(\mu)$ and
is given by
\begin{align}
		\mu_D&=\left[L_{\text{M}}^{1/2}(\mu)+\frac{(\mu-\gamma_1)}{r_1}\left(3L_{\text{M}}(\mu)+\kappa_n-2\frac{yq_c}{A_c}\right)\right]\mu_{N}\,.
	\end{align}
In contrast, the ``renormalized magnetic radius'' $\mu_D \langle r_{\text{M}1}^{2}\rangle$ is independent of $L_{\text{M}}^{1/2}(\mu)$ and the result for a \mbox{$J=1/2$} state reads
\begin{align}
  \mu_D \langle r_{\text{M}1}^{2}\rangle &= \frac{\mu_N}{\gamma_1 r_1} \left( \frac{3}{2}(1-y)^2 \kappa_n - y^2 \frac{yq_c}{A_c} \right)\,.
\end{align}

\section{Vector Current For Arbitrary Multipolarity}
\label{sec:AppGeneralized}
The generalized structure of the vector current for  arbitrary multipolarity can be deduced from \cref{eq:VectorFormFactor}. In particular, the electric form factors in the vector current with an even multipolarity $L$ for a particle with total spin $J$ appear as
	\begin{align}
		\left(i{\mu}_{\text{E}L}G_{\text{E}L}(q)\frac{1}{a(L)}\sqrt{\frac{4\pi}{2L+1}}q^{L}\sum_M Y^*_{LM}(\mb{e}_{\mb{q}})\left(\tilde{T}^{\left[LM\right]}_{J}\right)_{\beta'\beta}\right)
		\frac{\left(\mb{p}'+\mb{p}\right)^{k}}{2M_{nc}},
	\end{align}
where $a(L)$ denotes the leading denominators of the Legendre polynomials given by
	\begin{align}
		a(L)=2^{\left(\text{Floor}\left[\frac{L}{2}\right]+\text{Floor}\left[\frac{L}{4}\right]+\text{Floor}\left[\frac{L}{8}\right]
		+\text{Floor}\left[\frac{L}{16}\right]+...\right)},
	\end{align}
while ${\mu}_{\text{E}L}$ stands for the electric multipole moment and $\text{Floor}[x]$ gives the greatest integer less than or equal to $x$.

The magnetic form factors in the vector current with an odd multipolarity $L$ for a particle with total spin $J$ appear as
	\begin{align}
		i{\mu}_{\text{M}L}G_{\text{M}L}(q)\frac{1}{a(L)}\sqrt{\frac{4\pi}{2L+1}}q^{L}\sum_M \sqrt{2} C_{\scriptscriptstyle{(1k)(LM)}}^{\scriptscriptstyle{L(M+k)}} Y^{*}_{L(M+k)}
		(\mb{e}_{\mb{q}})\left(\tilde{T}^{\left[LM\right]}_{J}\right)_{\beta'\beta},
	\end{align}
where ${\mu}_{\text{M}L}$ denotes the magnetic multipole moment.

\section{\textit{P}-Wave Halo EFT Wave Function}
\label{sec:AppPWaveWF}
In this Section, we calculate the wave function for a $P$-wave bound state mentioned in \cref{sec:E1}.

For the derivation of the wave function, we use 
\begin{align}
\label{eq:PWaveWF1}
  G_{i'i} \sim \frac{\ket{\psi_{i'}}\bra{\psi_i}}{E+B_1} \text{ for } E \rightarrow -B_1\,,
\end{align}
where $G_{i'i}$ is the fully interacting Green's function, $B_1$ is the binding energy, and $\ket{\psi_i}$ denotes the corresponding $P$-wave bound state.

Furthermore, we use
\begin{align}
\label{eq:PWaveWF2}
  G_{i'i} = G^0 \delta_{i'i} + G^0 T_{i'i} G^0\,,
\end{align}
where $G^0$ and $T_{i'i}$ are the free Green's function and the $P$-wave T-matrix, respectively.

Since the free Green's function gives no contribution to the pole, we find from \cref{eq:PWaveWF1} and \cref{eq:PWaveWF2} 
\begin{equation}
\label{eq:PWaveWF3}
\begin{aligned}
  \lim\limits_{E \to -B_1} \frac{\braket{\bs{k}'|\psi_{i'}}\braket{\psi_i|\bs{k}}}{E+B_1} = \lim\limits_{E \to -B_1} \bra{\bs{k}'}G^0 T_{i'i} G^0 \ket{\bs{k}}\,.
\end{aligned}
\end{equation}

We consider
\begin{equation}
\begin{aligned}
  \lim\limits_{E \to -B_1} \bra{\bs{k}'}G^0 T_{i'i} G^0 \ket{\bs{k}} =& \lim\limits_{E \to -B_1} \frac{1}{E-\frac{k'^2}{2m_R}+i\epsilon} \braket{\bs{k}'|T_{i'i}|\bs{k}} \frac{1}{E-\frac{k^2}{2m_R}+i\epsilon}\\
  =& \frac{1}{B_1+\frac{k'^2}{2m_R}} \left( g_1^2 D_1(-B_1)k'_{i'} k_i^*  \right) \frac{1}{B_1+\frac{k^2}{2m_R}}\,,
\end{aligned}
\end{equation}
where we have used
\begin{align}
  \braket{\bs{k}'|T_{i'i}|\bs{k}} = g_1^2 D_1(E)k'_{i'} k_i^*\,,
\end{align}
with $D_1(E)$ denoting the full dimer propagator for the $P$-wave bound state.

Making use of the expansion of the full dimer propagator around the bound state energy $B_1$
\begin{align}
  \lim\limits_{E \to -B_1} D_1(E)= \frac{Z_{\pi}}{E+B_1}\,,
\end{align}
we obtain
\begin{align}
  \lim\limits_{E \to -B_1} \bra{\bs{k}'}G^0 T_{i'i} G^0 \ket{\bs{k}} =& \lim\limits_{E \to -B_1} \frac{2m_Rg_1\sqrt{Z_\pi}k'_{i'}}{\gamma_1^2+k'^2} \frac{1}{E+B_1} \frac{2m_Rg_1\sqrt{Z_\pi}k_i^*}{\gamma_1^2+k^2}\,.
\end{align}

Comparing this result to \cref{eq:PWaveWF3} yields the wave function in momentum space
\begin{align}
  \braket{\bs{k}|\psi_i} = \frac{2m_Rg_1 \sqrt{Z_\pi} k_i}{\gamma_1^2+k^2}\,.
\end{align}

Finally, coupling the orbital angular momentum with quantum numbers $(1i)$ with the spin of the neutron with quantum numbers $(\frac{1}{2}\alpha)$
to the total angular momentum $(\frac{3}{2}\beta)$
leads to
\begin{align}
  \braket{\bs{k}|\psi^{\alpha\beta}} = \frac{2m_Rg_1 \sqrt{Z_\pi} k_i C_{\scriptscriptstyle{(1i)(\frac{1}{2}\alpha)}}
										^{\scriptscriptstyle{\frac{3}{2}\beta}}}{\gamma_1^2+k^2}\,.
\end{align}

\newpage
% \bibliographystyle{apsrev4-1}
% \bibliography{literature}

\begin{thebibliography}{36}%
\makeatletter
\providecommand \@ifxundefined [1]{%
 \@ifx{#1\undefined}
}%
\providecommand \@ifnum [1]{%
 \ifnum #1\expandafter \@firstoftwo
 \else \expandafter \@secondoftwo
 \fi
}%
\providecommand \@ifx [1]{%
 \ifx #1\expandafter \@firstoftwo
 \else \expandafter \@secondoftwo
 \fi
}%
\providecommand \natexlab [1]{#1}%
\providecommand \enquote  [1]{``#1''}%
\providecommand \bibnamefont  [1]{#1}%
\providecommand \bibfnamefont [1]{#1}%
\providecommand \citenamefont [1]{#1}%
\providecommand \href@noop [0]{\@secondoftwo}%
\providecommand \href [0]{\begingroup \@sanitize@url \@href}%
\providecommand \@href[1]{\@@startlink{#1}\@@href}%
\providecommand \@@href[1]{\endgroup#1\@@endlink}%
\providecommand \@sanitize@url [0]{\catcode `\\12\catcode `\$12\catcode
  `\&12\catcode `\#12\catcode `\^12\catcode `\_12\catcode `\%12\relax}%
\providecommand \@@startlink[1]{}%
\providecommand \@@endlink[0]{}%
\providecommand \url  [0]{\begingroup\@sanitize@url \@url }%
\providecommand \@url [1]{\endgroup\@href {#1}{\urlprefix }}%
\providecommand \urlprefix  [0]{URL }%
\providecommand \Eprint [0]{\href }%
\providecommand \doibase [0]{http://dx.doi.org/}%
\providecommand \selectlanguage [0]{\@gobble}%
\providecommand \bibinfo  [0]{\@secondoftwo}%
\providecommand \bibfield  [0]{\@secondoftwo}%
\providecommand \translation [1]{[#1]}%
\providecommand \BibitemOpen [0]{}%
\providecommand \bibitemStop [0]{}%
\providecommand \bibitemNoStop [0]{.\EOS\space}%
\providecommand \EOS [0]{\spacefactor3000\relax}%
\providecommand \BibitemShut  [1]{\csname bibitem#1\endcsname}%
\let\auto@bib@innerbib\@empty
%</preamble>
\bibitem [{\citenamefont {Hansen}\ \emph {et~al.}(1995)\citenamefont {Hansen},
  \citenamefont {Jensen},\ and\ \citenamefont {Jonson}}]{Hansen:1995pu}%
  \BibitemOpen
  \bibfield  {author} {\bibinfo {author} {\bibfnamefont {P.~G.}\ \bibnamefont
  {Hansen}}, \bibinfo {author} {\bibfnamefont {A.~S.}\ \bibnamefont {Jensen}},
  \ and\ \bibinfo {author} {\bibfnamefont {B.}~\bibnamefont {Jonson}},\ }\href
  {\doibase 10.1146/annurev.ns.45.120195.003111} {\bibfield  {journal}
  {\bibinfo  {journal} {Ann. Rev. Nucl. Part. Sci.}\ }\textbf {\bibinfo
  {volume} {45}},\ \bibinfo {pages} {591} (\bibinfo {year} {1995})}\BibitemShut
  {NoStop}%
\bibitem [{\citenamefont {Jonson}(2004)}]{Jonson:2004}%
  \BibitemOpen
  \bibfield  {author} {\bibinfo {author} {\bibfnamefont {B.}~\bibnamefont
  {Jonson}},\ }\href {\doibase http://dx.doi.org/10.1016/j.physrep.2003.07.004}
  {\bibfield  {journal} {\bibinfo  {journal} {Physics Reports}\ }\textbf
  {\bibinfo {volume} {389}},\ \bibinfo {pages} {1 } (\bibinfo {year}
  {2004})}\BibitemShut {NoStop}%
\bibitem [{\citenamefont {Riisager}(2013)}]{Riisager:2012it}%
  \BibitemOpen
  \bibfield  {author} {\bibinfo {author} {\bibfnamefont {K.}~\bibnamefont
  {Riisager}},\ }\href {\doibase 10.1088/0031-8949/2013/T152/014001} {\bibfield
   {journal} {\bibinfo  {journal} {Phys. Scripta T}\ }\textbf {\bibinfo
  {volume} {152}},\ \bibinfo {pages} {014001} (\bibinfo {year} {2013})},\
  \Eprint {http://arxiv.org/abs/1208.6415} {arXiv:1208.6415 [nucl-ex]}
  \BibitemShut {NoStop}%
\bibitem [{\citenamefont {Tanihata}(2016)}]{Tanihata:2016zgp}%
  \BibitemOpen
  \bibfield  {author} {\bibinfo {author} {\bibfnamefont {I.}~\bibnamefont
  {Tanihata}},\ }\href {\doibase 10.1140/epjp/i2016-16090-x} {\bibfield
  {journal} {\bibinfo  {journal} {Eur. Phys. J. Plus}\ }\textbf {\bibinfo
  {volume} {131}},\ \bibinfo {pages} {90} (\bibinfo {year} {2016})}\BibitemShut
  {NoStop}%
\bibitem [{\citenamefont {Jensen}\ \emph {et~al.}(2004)\citenamefont {Jensen},
  \citenamefont {Riisager}, \citenamefont {Fedorov},\ and\ \citenamefont
  {Garrido}}]{Jensen:2004zz}%
  \BibitemOpen
  \bibfield  {author} {\bibinfo {author} {\bibfnamefont {A.~S.}\ \bibnamefont
  {Jensen}}, \bibinfo {author} {\bibfnamefont {K.}~\bibnamefont {Riisager}},
  \bibinfo {author} {\bibfnamefont {D.~V.}\ \bibnamefont {Fedorov}}, \ and\
  \bibinfo {author} {\bibfnamefont {E.}~\bibnamefont {Garrido}},\ }\href
  {\doibase 10.1103/RevModPhys.76.215} {\bibfield  {journal} {\bibinfo
  {journal} {Rev. Mod. Phys.}\ }\textbf {\bibinfo {volume} {76}},\ \bibinfo
  {pages} {215} (\bibinfo {year} {2004})}\BibitemShut {NoStop}%
\bibitem [{\citenamefont {Braaten}\ and\ \citenamefont
  {Hammer}(2006)}]{Braaten:2004rn}%
  \BibitemOpen
  \bibfield  {author} {\bibinfo {author} {\bibfnamefont {E.}~\bibnamefont
  {Braaten}}\ and\ \bibinfo {author} {\bibfnamefont {H.-W.}\ \bibnamefont
  {Hammer}},\ }\href {\doibase 10.1016/j.physrep.2006.03.001} {\bibfield
  {journal} {\bibinfo  {journal} {Phys. Rept.}\ }\textbf {\bibinfo {volume}
  {428}},\ \bibinfo {pages} {259} (\bibinfo {year} {2006})},\ \Eprint
  {http://arxiv.org/abs/cond-mat/0410417} {arXiv:cond-mat/0410417} \BibitemShut
  {NoStop}%
\bibitem [{\citenamefont {Hammer}\ \emph {et~al.}(2017)\citenamefont {Hammer},
  \citenamefont {Ji},\ and\ \citenamefont {Phillips}}]{Hammer:2017tjm}%
  \BibitemOpen
  \bibfield  {author} {\bibinfo {author} {\bibfnamefont {H.-W.}\ \bibnamefont
  {Hammer}}, \bibinfo {author} {\bibfnamefont {C.}~\bibnamefont {Ji}}, \ and\
  \bibinfo {author} {\bibfnamefont {D.~R.}\ \bibnamefont {Phillips}},\ }\href
  {\doibase 10.1088/1361-6471/aa83db} {\bibfield  {journal} {\bibinfo
  {journal} {J. Phys. G}\ }\textbf {\bibinfo {volume} {44}},\ \bibinfo {pages}
  {103002} (\bibinfo {year} {2017})},\ \Eprint
  {http://arxiv.org/abs/1702.08605} {arXiv:1702.08605 [nucl-th]} \BibitemShut
  {NoStop}%
\bibitem [{\citenamefont {Bertulani}\ \emph {et~al.}(2002)\citenamefont
  {Bertulani}, \citenamefont {Hammer},\ and\ \citenamefont
  {Van~Kolck}}]{Bertulani:2002sz}%
  \BibitemOpen
  \bibfield  {author} {\bibinfo {author} {\bibfnamefont {C.~A.}\ \bibnamefont
  {Bertulani}}, \bibinfo {author} {\bibfnamefont {H.-W.}\ \bibnamefont
  {Hammer}}, \ and\ \bibinfo {author} {\bibfnamefont {U.}~\bibnamefont
  {Van~Kolck}},\ }\href {\doibase 10.1016/S0375-9474(02)01270-8} {\bibfield
  {journal} {\bibinfo  {journal} {Nucl. Phys. A}\ }\textbf {\bibinfo {volume}
  {712}},\ \bibinfo {pages} {37} (\bibinfo {year} {2002})},\ \Eprint
  {http://arxiv.org/abs/nucl-th/0205063} {arXiv:nucl-th/0205063} \BibitemShut
  {NoStop}%
\bibitem [{\citenamefont {Bedaque}\ \emph {et~al.}(2003)\citenamefont
  {Bedaque}, \citenamefont {Hammer},\ and\ \citenamefont {van
  Kolck}}]{Bedaque:2003wa}%
  \BibitemOpen
  \bibfield  {author} {\bibinfo {author} {\bibfnamefont {P.~F.}\ \bibnamefont
  {Bedaque}}, \bibinfo {author} {\bibfnamefont {H.-W.}\ \bibnamefont {Hammer}},
  \ and\ \bibinfo {author} {\bibfnamefont {U.}~\bibnamefont {van Kolck}},\
  }\href {\doibase 10.1016/j.physletb.2003.07.049} {\bibfield  {journal}
  {\bibinfo  {journal} {Phys. Lett. B}\ }\textbf {\bibinfo {volume} {569}},\
  \bibinfo {pages} {159} (\bibinfo {year} {2003})},\ \Eprint
  {http://arxiv.org/abs/nucl-th/0304007} {arXiv:nucl-th/0304007} \BibitemShut
  {NoStop}%
\bibitem [{\citenamefont {Nakamura}\ \emph {et~al.}(2009)\citenamefont
  {Nakamura}, \citenamefont {Kobayashi}, \citenamefont {Kondo}, \citenamefont
  {Satou}, \citenamefont {Aoi}, \citenamefont {Baba}, \citenamefont {Deguchi},
  \citenamefont {Fukuda}, \citenamefont {Gibelin}, \citenamefont {Inabe},
  \citenamefont {Ishihara}, \citenamefont {Kameda}, \citenamefont {Kawada},
  \citenamefont {Kubo}, \citenamefont {Kusaka}, \citenamefont {Mengoni},
  \citenamefont {Motobayashi}, \citenamefont {Ohnishi}, \citenamefont {Ohtake},
  \citenamefont {Orr}, \citenamefont {Otsu}, \citenamefont {Otsuka},
  \citenamefont {Saito}, \citenamefont {Sakurai}, \citenamefont {Shimoura},
  \citenamefont {Sumikama}, \citenamefont {Takeda}, \citenamefont {Takeshita},
  \citenamefont {Takechi}, \citenamefont {Takeuchi}, \citenamefont {Tanaka},
  \citenamefont {Tanaka}, \citenamefont {Tanaka}, \citenamefont {Togano},
  \citenamefont {Utsuno}, \citenamefont {Yoneda}, \citenamefont {Yoshida},\
  and\ \citenamefont {Yoshida}}]{PhysRevLett.103.262501}%
  \BibitemOpen
  \bibfield  {author} {\bibinfo {author} {\bibfnamefont {T.}~\bibnamefont
  {Nakamura}}, \bibinfo {author} {\bibfnamefont {N.}~\bibnamefont {Kobayashi}},
  \bibinfo {author} {\bibfnamefont {Y.}~\bibnamefont {Kondo}}, \bibinfo
  {author} {\bibfnamefont {Y.}~\bibnamefont {Satou}}, \bibinfo {author}
  {\bibfnamefont {N.}~\bibnamefont {Aoi}}, \bibinfo {author} {\bibfnamefont
  {H.}~\bibnamefont {Baba}}, \bibinfo {author} {\bibfnamefont {S.}~\bibnamefont
  {Deguchi}}, \bibinfo {author} {\bibfnamefont {N.}~\bibnamefont {Fukuda}},
  \bibinfo {author} {\bibfnamefont {J.}~\bibnamefont {Gibelin}}, \bibinfo
  {author} {\bibfnamefont {N.}~\bibnamefont {Inabe}}, \bibinfo {author}
  {\bibfnamefont {M.}~\bibnamefont {Ishihara}}, \bibinfo {author}
  {\bibfnamefont {D.}~\bibnamefont {Kameda}}, \bibinfo {author} {\bibfnamefont
  {Y.}~\bibnamefont {Kawada}}, \bibinfo {author} {\bibfnamefont
  {T.}~\bibnamefont {Kubo}}, \bibinfo {author} {\bibfnamefont {K.}~\bibnamefont
  {Kusaka}}, \bibinfo {author} {\bibfnamefont {A.}~\bibnamefont {Mengoni}},
  \bibinfo {author} {\bibfnamefont {T.}~\bibnamefont {Motobayashi}}, \bibinfo
  {author} {\bibfnamefont {T.}~\bibnamefont {Ohnishi}}, \bibinfo {author}
  {\bibfnamefont {M.}~\bibnamefont {Ohtake}}, \bibinfo {author} {\bibfnamefont
  {N.~A.}\ \bibnamefont {Orr}}, \bibinfo {author} {\bibfnamefont
  {H.}~\bibnamefont {Otsu}}, \bibinfo {author} {\bibfnamefont {T.}~\bibnamefont
  {Otsuka}}, \bibinfo {author} {\bibfnamefont {A.}~\bibnamefont {Saito}},
  \bibinfo {author} {\bibfnamefont {H.}~\bibnamefont {Sakurai}}, \bibinfo
  {author} {\bibfnamefont {S.}~\bibnamefont {Shimoura}}, \bibinfo {author}
  {\bibfnamefont {T.}~\bibnamefont {Sumikama}}, \bibinfo {author}
  {\bibfnamefont {H.}~\bibnamefont {Takeda}}, \bibinfo {author} {\bibfnamefont
  {E.}~\bibnamefont {Takeshita}}, \bibinfo {author} {\bibfnamefont
  {M.}~\bibnamefont {Takechi}}, \bibinfo {author} {\bibfnamefont
  {S.}~\bibnamefont {Takeuchi}}, \bibinfo {author} {\bibfnamefont
  {K.}~\bibnamefont {Tanaka}}, \bibinfo {author} {\bibfnamefont {K.~N.}\
  \bibnamefont {Tanaka}}, \bibinfo {author} {\bibfnamefont {N.}~\bibnamefont
  {Tanaka}}, \bibinfo {author} {\bibfnamefont {Y.}~\bibnamefont {Togano}},
  \bibinfo {author} {\bibfnamefont {Y.}~\bibnamefont {Utsuno}}, \bibinfo
  {author} {\bibfnamefont {K.}~\bibnamefont {Yoneda}}, \bibinfo {author}
  {\bibfnamefont {A.}~\bibnamefont {Yoshida}}, \ and\ \bibinfo {author}
  {\bibfnamefont {K.}~\bibnamefont {Yoshida}},\ }\href {\doibase
  10.1103/PhysRevLett.103.262501} {\bibfield  {journal} {\bibinfo  {journal}
  {Phys. Rev. Lett.}\ }\textbf {\bibinfo {volume} {103}},\ \bibinfo {pages}
  {262501} (\bibinfo {year} {2009})}\BibitemShut {NoStop}%
\bibitem [{\citenamefont {Nakamura}\ \emph {et~al.}(2014)\citenamefont
  {Nakamura} \emph {et~al.}}]{Nakamura:2014hxa}%
  \BibitemOpen
  \bibfield  {author} {\bibinfo {author} {\bibfnamefont {T.}~\bibnamefont
  {Nakamura}} \emph {et~al.},\ }\href {\doibase 10.1103/PhysRevLett.112.142501}
  {\bibfield  {journal} {\bibinfo  {journal} {Phys. Rev. Lett.}\ }\textbf
  {\bibinfo {volume} {112}},\ \bibinfo {pages} {142501} (\bibinfo {year}
  {2014})}\BibitemShut {NoStop}%
\bibitem [{\citenamefont {Kobayashi}\ \emph {et~al.}(2014)\citenamefont
  {Kobayashi} \emph {et~al.}}]{Kobayashi:2014owa}%
  \BibitemOpen
  \bibfield  {author} {\bibinfo {author} {\bibfnamefont {N.}~\bibnamefont
  {Kobayashi}} \emph {et~al.},\ }\href {\doibase
  10.1103/PhysRevLett.112.242501} {\bibfield  {journal} {\bibinfo  {journal}
  {Phys. Rev. Lett.}\ }\textbf {\bibinfo {volume} {112}},\ \bibinfo {pages}
  {242501} (\bibinfo {year} {2014})}\BibitemShut {NoStop}%
\bibitem [{\citenamefont {Zhong-Zhou}\ \emph {et~al.}(2001)\citenamefont
  {Zhong-Zhou}, \citenamefont {Bao-Qiu}, \citenamefont {Zhong-Yu},\ and\
  \citenamefont {Gong-Ou}}]{Zhong_Zhou_2001}%
  \BibitemOpen
  \bibfield  {author} {\bibinfo {author} {\bibfnamefont {R.}~\bibnamefont
  {Zhong-Zhou}}, \bibinfo {author} {\bibfnamefont {C.}~\bibnamefont {Bao-Qiu}},
  \bibinfo {author} {\bibfnamefont {M.}~\bibnamefont {Zhong-Yu}}, \ and\
  \bibinfo {author} {\bibfnamefont {X.}~\bibnamefont {Gong-Ou}},\ }\href
  {\doibase 10.1088/0253-6102/35/6/717} {\bibfield  {journal} {\bibinfo
  {journal} {Communications in Theoretical Physics}\ }\textbf {\bibinfo
  {volume} {35}},\ \bibinfo {pages} {717} (\bibinfo {year} {2001})}\BibitemShut
  {NoStop}%
\bibitem [{\citenamefont {Hamamoto}(2010)}]{PhysRevC.81.021304}%
  \BibitemOpen
  \bibfield  {author} {\bibinfo {author} {\bibfnamefont {I.}~\bibnamefont
  {Hamamoto}},\ }\href {\doibase 10.1103/PhysRevC.81.021304} {\bibfield
  {journal} {\bibinfo  {journal} {Phys. Rev. C}\ }\textbf {\bibinfo {volume}
  {81}},\ \bibinfo {pages} {021304} (\bibinfo {year} {2010})}\BibitemShut
  {NoStop}%
\bibitem [{\citenamefont {Urata}\ \emph {et~al.}(2011)\citenamefont {Urata},
  \citenamefont {Hagino},\ and\ \citenamefont {Sagawa}}]{PhysRevC.83.041303}%
  \BibitemOpen
  \bibfield  {author} {\bibinfo {author} {\bibfnamefont {Y.}~\bibnamefont
  {Urata}}, \bibinfo {author} {\bibfnamefont {K.}~\bibnamefont {Hagino}}, \
  and\ \bibinfo {author} {\bibfnamefont {H.}~\bibnamefont {Sagawa}},\ }\href
  {\doibase 10.1103/PhysRevC.83.041303} {\bibfield  {journal} {\bibinfo
  {journal} {Phys. Rev. C}\ }\textbf {\bibinfo {volume} {83}},\ \bibinfo
  {pages} {041303} (\bibinfo {year} {2011})}\BibitemShut {NoStop}%
\bibitem [{\citenamefont {Minomo}\ \emph {et~al.}(2012)\citenamefont {Minomo},
  \citenamefont {Sumi}, \citenamefont {Kimura}, \citenamefont {Ogata},
  \citenamefont {Shimizu},\ and\ \citenamefont
  {Yahiro}}]{PhysRevLett.108.052503}%
  \BibitemOpen
  \bibfield  {author} {\bibinfo {author} {\bibfnamefont {K.}~\bibnamefont
  {Minomo}}, \bibinfo {author} {\bibfnamefont {T.}~\bibnamefont {Sumi}},
  \bibinfo {author} {\bibfnamefont {M.}~\bibnamefont {Kimura}}, \bibinfo
  {author} {\bibfnamefont {K.}~\bibnamefont {Ogata}}, \bibinfo {author}
  {\bibfnamefont {Y.~R.}\ \bibnamefont {Shimizu}}, \ and\ \bibinfo {author}
  {\bibfnamefont {M.}~\bibnamefont {Yahiro}},\ }\href {\doibase
  10.1103/PhysRevLett.108.052503} {\bibfield  {journal} {\bibinfo  {journal}
  {Phys. Rev. Lett.}\ }\textbf {\bibinfo {volume} {108}},\ \bibinfo {pages}
  {052503} (\bibinfo {year} {2012})}\BibitemShut {NoStop}%
\bibitem [{\citenamefont {Shubhchintak}\ and\ \citenamefont
  {Chatterjee}(2014)}]{SHUBHCHINTAK201499}%
  \BibitemOpen
  \bibfield  {author} {\bibinfo {author} {\bibnamefont {Shubhchintak}}\ and\
  \bibinfo {author} {\bibfnamefont {R.}~\bibnamefont {Chatterjee}},\ }\href
  {\doibase https://doi.org/10.1016/j.nuclphysa.2013.11.010} {\bibfield
  {journal} {\bibinfo  {journal} {Nucl. Phys. A}\ }\textbf {\bibinfo {volume}
  {922}},\ \bibinfo {pages} {99} (\bibinfo {year} {2014})}\BibitemShut
  {NoStop}%
\bibitem [{\citenamefont {Hong}\ \emph {et~al.}(2017)\citenamefont {Hong},
  \citenamefont {Bertulani},\ and\ \citenamefont {Kruppa}}]{Hong:2017vul}%
  \BibitemOpen
  \bibfield  {author} {\bibinfo {author} {\bibfnamefont {J.}~\bibnamefont
  {Hong}}, \bibinfo {author} {\bibfnamefont {C.~A.}\ \bibnamefont {Bertulani}},
  \ and\ \bibinfo {author} {\bibfnamefont {A.~T.}\ \bibnamefont {Kruppa}},\
  }\href {\doibase 10.1103/PhysRevC.96.064603} {\bibfield  {journal} {\bibinfo
  {journal} {Phys. Rev. C}\ }\textbf {\bibinfo {volume} {96}},\ \bibinfo
  {pages} {064603} (\bibinfo {year} {2017})},\ \Eprint
  {http://arxiv.org/abs/1712.00639} {arXiv:1712.00639 [nucl-th]} \BibitemShut
  {NoStop}%
\bibitem [{\citenamefont {Li}\ \emph {et~al.}(2022)\citenamefont {Li},
  \citenamefont {Michel}, \citenamefont {Li},\ and\ \citenamefont
  {Zuo}}]{Li:2022qzo}%
  \BibitemOpen
  \bibfield  {author} {\bibinfo {author} {\bibfnamefont {J.~G.}\ \bibnamefont
  {Li}}, \bibinfo {author} {\bibfnamefont {N.}~\bibnamefont {Michel}}, \bibinfo
  {author} {\bibfnamefont {H.~H.}\ \bibnamefont {Li}}, \ and\ \bibinfo {author}
  {\bibfnamefont {W.}~\bibnamefont {Zuo}},\ }\href {\doibase
  10.1016/j.physletb.2022.137225} {\bibfield  {journal} {\bibinfo  {journal}
  {Phys. Lett. B}\ }\textbf {\bibinfo {volume} {832}},\ \bibinfo {pages}
  {137225} (\bibinfo {year} {2022})},\ \Eprint
  {http://arxiv.org/abs/2206.03765} {arXiv:2206.03765 [nucl-th]} \BibitemShut
  {NoStop}%
\bibitem [{\citenamefont {Hammer}\ and\ \citenamefont
  {Phillips}(2011)}]{Hammer:2011ye}%
  \BibitemOpen
  \bibfield  {author} {\bibinfo {author} {\bibfnamefont {H.-W.}\ \bibnamefont
  {Hammer}}\ and\ \bibinfo {author} {\bibfnamefont {D.~R.}\ \bibnamefont
  {Phillips}},\ }\href {\doibase 10.1016/j.nuclphysa.2011.06.028} {\bibfield
  {journal} {\bibinfo  {journal} {Nucl. Phys. A}\ }\textbf {\bibinfo {volume}
  {865}},\ \bibinfo {pages} {17} (\bibinfo {year} {2011})},\ \Eprint
  {http://arxiv.org/abs/1103.1087} {arXiv:1103.1087 [nucl-th]} \BibitemShut
  {NoStop}%
\bibitem [{\citenamefont {He}(2019)}]{FBS60.16}%
  \BibitemOpen
  \bibfield  {author} {\bibinfo {author} {\bibfnamefont {F.}~\bibnamefont
  {He}},\ }\href@noop {} {\bibfield  {journal} {\bibinfo  {journal} {Few-Body
  Systems}\ }\textbf {\bibinfo {volume} {60}},\ \bibinfo {pages} {16} (\bibinfo
  {year} {2019})}\BibitemShut {NoStop}%
\bibitem [{\citenamefont {Kaplan}\ \emph
  {et~al.}(1998{\natexlab{a}})\citenamefont {Kaplan}, \citenamefont {Savage},\
  and\ \citenamefont {Wise}}]{Kaplan:1998tg}%
  \BibitemOpen
  \bibfield  {author} {\bibinfo {author} {\bibfnamefont {D.~B.}\ \bibnamefont
  {Kaplan}}, \bibinfo {author} {\bibfnamefont {M.~J.}\ \bibnamefont {Savage}},
  \ and\ \bibinfo {author} {\bibfnamefont {M.~B.}\ \bibnamefont {Wise}},\
  }\href {\doibase 10.1016/S0370-2693(98)00210-X} {\bibfield  {journal}
  {\bibinfo  {journal} {Phys. Lett. B}\ }\textbf {\bibinfo {volume} {424}},\
  \bibinfo {pages} {390} (\bibinfo {year} {1998}{\natexlab{a}})},\ \Eprint
  {http://arxiv.org/abs/nucl-th/9801034} {arXiv:nucl-th/9801034} \BibitemShut
  {NoStop}%
\bibitem [{\citenamefont {Kaplan}\ \emph
  {et~al.}(1998{\natexlab{b}})\citenamefont {Kaplan}, \citenamefont {Savage},\
  and\ \citenamefont {Wise}}]{Kaplan:1998we}%
  \BibitemOpen
  \bibfield  {author} {\bibinfo {author} {\bibfnamefont {D.~B.}\ \bibnamefont
  {Kaplan}}, \bibinfo {author} {\bibfnamefont {M.~J.}\ \bibnamefont {Savage}},
  \ and\ \bibinfo {author} {\bibfnamefont {M.~B.}\ \bibnamefont {Wise}},\
  }\href {\doibase 10.1016/S0550-3213(98)00440-4} {\bibfield  {journal}
  {\bibinfo  {journal} {Nucl. Phys. B}\ }\textbf {\bibinfo {volume} {534}},\
  \bibinfo {pages} {329} (\bibinfo {year} {1998}{\natexlab{b}})},\ \Eprint
  {http://arxiv.org/abs/nucl-th/9802075} {arXiv:nucl-th/9802075} \BibitemShut
  {NoStop}%
\bibitem [{\citenamefont {Hammer}\ and\ \citenamefont
  {Lee}(2009)}]{Hammer:2009zh}%
  \BibitemOpen
  \bibfield  {author} {\bibinfo {author} {\bibfnamefont {H.~W.}\ \bibnamefont
  {Hammer}}\ and\ \bibinfo {author} {\bibfnamefont {D.}~\bibnamefont {Lee}},\
  }\href {\doibase 10.1016/j.physletb.2009.10.033} {\bibfield  {journal}
  {\bibinfo  {journal} {Phys. Lett. B}\ }\textbf {\bibinfo {volume} {681}},\
  \bibinfo {pages} {500} (\bibinfo {year} {2009})},\ \Eprint
  {http://arxiv.org/abs/0907.1763} {arXiv:0907.1763 [nucl-th]} \BibitemShut
  {NoStop}%
\bibitem [{\citenamefont {Hammer}\ and\ \citenamefont
  {Lee}(2010)}]{Hammer:2010fw}%
  \BibitemOpen
  \bibfield  {author} {\bibinfo {author} {\bibfnamefont {H.~W.}\ \bibnamefont
  {Hammer}}\ and\ \bibinfo {author} {\bibfnamefont {D.}~\bibnamefont {Lee}},\
  }\href {\doibase 10.1016/j.aop.2010.06.006} {\bibfield  {journal} {\bibinfo
  {journal} {Annals Phys.}\ }\textbf {\bibinfo {volume} {325}},\ \bibinfo
  {pages} {2212} (\bibinfo {year} {2010})},\ \Eprint
  {http://arxiv.org/abs/1002.4603} {arXiv:1002.4603 [nucl-th]} \BibitemShut
  {NoStop}%
\bibitem [{\citenamefont
  {Shamsuzzoha~Basunia}(2010)}]{ShamsuzzohaBasunia:2010ggy}%
  \BibitemOpen
  \bibfield  {author} {\bibinfo {author} {\bibfnamefont {M.}~\bibnamefont
  {Shamsuzzoha~Basunia}},\ }\href {\doibase 10.1016/j.nds.2010.09.001}
  {\bibfield  {journal} {\bibinfo  {journal} {Nucl. Data Sheets}\ }\textbf
  {\bibinfo {volume} {111}},\ \bibinfo {pages} {2331} (\bibinfo {year}
  {2010})}\BibitemShut {NoStop}%
\bibitem [{\citenamefont {Khersonskii}\ \emph {et~al.}(1988)\citenamefont
  {Khersonskii}, \citenamefont {Moskalev},\ and\ \citenamefont
  {Varshalovich}}]{Khersonskii:1988krb}%
  \BibitemOpen
  \bibfield  {author} {\bibinfo {author} {\bibfnamefont {V.~K.}\ \bibnamefont
  {Khersonskii}}, \bibinfo {author} {\bibfnamefont {A.~N.}\ \bibnamefont
  {Moskalev}}, \ and\ \bibinfo {author} {\bibfnamefont {D.~A.}\ \bibnamefont
  {Varshalovich}},\ }\href {\doibase 10.1142/0270} {\emph {\bibinfo {title}
  {{Quantum Theory Of Angular Momentum}}}}\ (\bibinfo  {publisher} {World
  Scientific Publishing Company},\ \bibinfo {year} {1988})\BibitemShut
  {NoStop}%
\bibitem [{\citenamefont {Fernando}\ \emph {et~al.}(2015)\citenamefont
  {Fernando}, \citenamefont {Vaghani},\ and\ \citenamefont
  {Rupak}}]{Fernando:2015jyd}%
  \BibitemOpen
  \bibfield  {author} {\bibinfo {author} {\bibfnamefont {L.}~\bibnamefont
  {Fernando}}, \bibinfo {author} {\bibfnamefont {A.}~\bibnamefont {Vaghani}}, \
  and\ \bibinfo {author} {\bibfnamefont {G.}~\bibnamefont {Rupak}},\
  }\href@noop {} {\  (\bibinfo {year} {2015})},\ \Eprint
  {http://arxiv.org/abs/1511.04054} {arXiv:1511.04054 [nucl-th]} \BibitemShut
  {NoStop}%
\bibitem [{\citenamefont {Zelevinsky}\ and\ \citenamefont
  {Volya}(2017)}]{Zelevinskych12}%
  \BibitemOpen
  \bibfield  {author} {\bibinfo {author} {\bibfnamefont {V.}~\bibnamefont
  {Zelevinsky}}\ and\ \bibinfo {author} {\bibfnamefont {A.}~\bibnamefont
  {Volya}},\ }\enquote {\bibinfo {title} {Nuclear deformation},}\ in\
  \href@noop {} {\emph {\bibinfo {booktitle} {Physics of Atomic Nuclei}}}\
  (\bibinfo  {publisher} {John Wiley \& Sons, Ltd},\ \bibinfo {year} {2017})\
  Chap.~\bibinfo {chapter} {12}, pp.\ \bibinfo {pages} {223--250}\BibitemShut
  {NoStop}%
\bibitem [{\citenamefont {Greiner}\ \emph {et~al.}(1996)\citenamefont
  {Greiner}, \citenamefont {Bromley},\ and\ \citenamefont
  {Maruhn}}]{greiner1996nuclear}%
  \BibitemOpen
  \bibfield  {author} {\bibinfo {author} {\bibfnamefont {W.}~\bibnamefont
  {Greiner}}, \bibinfo {author} {\bibfnamefont {D.~A.}\ \bibnamefont
  {Bromley}}, \ and\ \bibinfo {author} {\bibfnamefont {J.}~\bibnamefont
  {Maruhn}},\ }\href@noop {} {\emph {\bibinfo {title} {Nuclear Models}}}\
  (\bibinfo  {publisher} {Springer Berlin Heidelberg},\ \bibinfo {year}
  {1996})\BibitemShut {NoStop}%
\bibitem [{\citenamefont {Greiner}\ and\ \citenamefont
  {Maruhn}(1996)}]{GreinerMaruhn}%
  \BibitemOpen
  \bibfield  {author} {\bibinfo {author} {\bibfnamefont {W.}~\bibnamefont
  {Greiner}}\ and\ \bibinfo {author} {\bibfnamefont {J.~A.}\ \bibnamefont
  {Maruhn}},\ }\href@noop {} {\emph {\bibinfo {title} {Nuclear Models}}}\
  (\bibinfo  {publisher} {Springer},\ \bibinfo {year} {1996})\BibitemShut
  {NoStop}%
\bibitem [{\citenamefont {Typel}\ and\ \citenamefont
  {Baur}(2005)}]{Typel:2004us}%
  \BibitemOpen
  \bibfield  {author} {\bibinfo {author} {\bibfnamefont {S.}~\bibnamefont
  {Typel}}\ and\ \bibinfo {author} {\bibfnamefont {G.}~\bibnamefont {Baur}},\
  }\href {\doibase 10.1016/j.nuclphysa.2005.05.145} {\bibfield  {journal}
  {\bibinfo  {journal} {Nucl. Phys. A}\ }\textbf {\bibinfo {volume} {759}},\
  \bibinfo {pages} {247} (\bibinfo {year} {2005})},\ \Eprint
  {http://arxiv.org/abs/nucl-th/0411069} {arXiv:nucl-th/0411069} \BibitemShut
  {NoStop}%
\bibitem [{\citenamefont {Bertulani}(2009)}]{Bertulani:2009zk}%
  \BibitemOpen
  \bibfield  {author} {\bibinfo {author} {\bibfnamefont {C.~A.}\ \bibnamefont
  {Bertulani}},\ }\href@noop {} {\  (\bibinfo {year} {2009})},\ \Eprint
  {http://arxiv.org/abs/0908.4307} {arXiv:0908.4307 [nucl-th]} \BibitemShut
  {NoStop}%
\bibitem [{\citenamefont {Ryberg}\ \emph {et~al.}(2020)\citenamefont {Ryberg},
  \citenamefont {Forss\'en}, \citenamefont {Phillips},\ and\ \citenamefont {van
  Kolck}}]{Ryberg:2019cvj}%
  \BibitemOpen
  \bibfield  {author} {\bibinfo {author} {\bibfnamefont {E.}~\bibnamefont
  {Ryberg}}, \bibinfo {author} {\bibfnamefont {C.}~\bibnamefont {Forss\'en}},
  \bibinfo {author} {\bibfnamefont {D.~R.}\ \bibnamefont {Phillips}}, \ and\
  \bibinfo {author} {\bibfnamefont {U.}~\bibnamefont {van Kolck}},\ }\href
  {\doibase 10.1140/epja/s10050-019-00001-1} {\bibfield  {journal} {\bibinfo
  {journal} {Eur. Phys. J. A}\ }\textbf {\bibinfo {volume} {56}},\ \bibinfo
  {pages} {7} (\bibinfo {year} {2020})},\ \Eprint
  {http://arxiv.org/abs/1905.01107} {arXiv:1905.01107 [nucl-th]} \BibitemShut
  {NoStop}%
\bibitem [{\citenamefont {Papenbrock}\ and\ \citenamefont
  {Weidenm\"uller}(2020)}]{Papenbrock:2020zhh}%
  \BibitemOpen
  \bibfield  {author} {\bibinfo {author} {\bibfnamefont {T.}~\bibnamefont
  {Papenbrock}}\ and\ \bibinfo {author} {\bibfnamefont {H.~A.}\ \bibnamefont
  {Weidenm\"uller}},\ }\href {\doibase 10.1103/PhysRevC.102.044324} {\bibfield
  {journal} {\bibinfo  {journal} {Phys. Rev. C}\ }\textbf {\bibinfo {volume}
  {102}},\ \bibinfo {pages} {044324} (\bibinfo {year} {2020})},\ \Eprint
  {http://arxiv.org/abs/2005.11865} {arXiv:2005.11865 [nucl-th]} \BibitemShut
  {NoStop}%
\bibitem [{\citenamefont {Alnamlah}\ \emph {et~al.}(2021)\citenamefont
  {Alnamlah}, \citenamefont {P\'erez},\ and\ \citenamefont
  {Phillips}}]{Alnamlah:2020cko}%
  \BibitemOpen
  \bibfield  {author} {\bibinfo {author} {\bibfnamefont {I.~K.}\ \bibnamefont
  {Alnamlah}}, \bibinfo {author} {\bibfnamefont {E.~A.~C.}\ \bibnamefont
  {P\'erez}}, \ and\ \bibinfo {author} {\bibfnamefont {D.~R.}\ \bibnamefont
  {Phillips}},\ }\href {\doibase 10.1103/PhysRevC.104.064311} {\bibfield
  {journal} {\bibinfo  {journal} {Phys. Rev. C}\ }\textbf {\bibinfo {volume}
  {104}},\ \bibinfo {pages} {064311} (\bibinfo {year} {2021})},\ \Eprint
  {http://arxiv.org/abs/2011.01083} {arXiv:2011.01083 [nucl-th]} \BibitemShut
  {NoStop}%
\end{thebibliography}
%

\end{document}